\documentclass[a4paper,11pt]{article}
\usepackage{jheppub} 
\usepackage{graphicx}
\usepackage{amsmath, amssymb, amsthm, mathtools}
\usepackage{tikz}
\usepackage{tikz-network}
\usetikzlibrary{patterns}

\usepackage{bm}
\usepackage{multirow}
\usepackage{braket}
\usepackage{float}

\title{Non-invertible symmetry and vertex operator algebra outer-automorphism}

\author[a,b]{Kazunobu Maruyoshi}
\author[c]{, Hyejung Moon}
\author[c,d]{, Jaewon Song}
\affiliation[a]{Faculty of Science and Technology, Seikei University\\
3-3-1 Kichijoji-Kitamachi, Musashino-shi, Tokyo, 180-8633, Japan}
\affiliation[b]{Nambu Yoichiro Institute of Theoretical and Experimental Physics\\ Osaka Metropolitan University, 3-3-138, Sugimoto, Sumiyoshi-ku, Osaka, 558-8585, Japan}
\affiliation[c]{Department of Physics, Korea Advanced Institute of Science and Technology\\ 291 Daehak-ro, Yuseong-gu, Daejeon 34141, Republic of Korea}
\affiliation[d]{Walter Burke Institute for Theoretical Physics \& Leinweber Forum for Theoretical Physics\\California Institute of Technology, 
Pasadena, CA 91125, USA}

\emailAdd{maruyoshi@st.seikei.ac.jp}
\emailAdd{moon130@kaist.ac.kr}
\emailAdd{jaewon.song@kaist.ac.kr}

\preprint{CALT-TH 2026-030}

\abstract{We argue that the non-invertible symmetry of $\mathcal{N}=4$ super Yang-Mills theory, obtained by combining S-duality, half-space gauging, and R-symmetry twist, is realized as an outer-automorphism of the associated vertex operator algebra. Hence, the Schur and Macdonald index of $\mathcal{N}=4$ super Yang-Mills theory with non-invertible symmetry twist corresponds to the (refined) vacuum character of the associated vertex operator algebra with the outer-automorphism twist. We provide a few checks of our proposal from the large-$N$ limit and by matching the fusion rules, which involve charge conjugation on local operators. The `charged' Schur index indeed matches the outer-automorphism-twisted vacuum character. 
}

\begin{document}
\maketitle
\flushbottom

\section{Introduction}\label{sec:intro}

In recent years, there have been rapid developments in expanding the concept of symmetry \cite{Gaiotto:2014kfa}. The main idea was that the symmetry charge operator can be thought of as a topological defect (of co-dimension 1) in spacetime. The usual group multiplication is realized as a fusion rule for the topological defect operator. This viewpoint allows generalization of the usual symmetry concept by choosing the topological defect to have higher co-dimensions. Moreover, if we accept that all topological defects are symmetry generators, we can even come up with the concept of `non-invertible' symmetry where the algebra of the defect does not form a group, instead a fusion category \cite{Chang:2018iay}. 

In four-dimensional field theory, the first examples of non-invertible symmetry (NIS) were realized by \cite{Choi:2021kmx, Kaidi:2021xfk}, which can be considered as a 4-dimensional uplift of $2d$ Kramers-Wannier duality \cite{Chang:2018iay}. These non-invertible symmetries can be realized whenever a theory can be dualized to itself. We focus on $\mathcal{N}=4$ super Yang-Mills (SYM) theory with gauge group $G=SU(N)$, which has well-known S-duality. The topological defect operator corresponding to the non-invertible symmetry is formed by a combination of S-duality domain-wall and half-space gauging of 1-form symmetry. At the self-dual points, such as $\tau_{{\rm YM}}=i$ (where $\tau_{{\rm YM}}$ is the usual complexified gauge coupling), this domain wall becomes an operator of the same theory. Since S-duality acts non-trivially on supercharges, the duality non-invertible symmetry does not preserve supersymmetry. To partially preserve the supersymmetry, we need to attach an additional R-symmetry rotation \cite{Kaidi:2022uux}. We will focus on these supersymmetrized versions of the NIS. See \cite{Kaidi:2022uux,Choi:2022zal,Kaidi:2023maf} for various aspects of NIS in $\mathcal{N}=4$ SYM theory, and \cite{Shao:2023gho, Schafer-Nameki:2023jdn} for reviews of NIS and references therein.

Any $4d$ $\mathcal{N}=2$ superconformal field theory has a subsector described by a vertex operator algebra \cite{Beem:2013sza}. In the case of $\mathcal{N}=4$ SYM theory, the associated VOA is given by a W-algebra with $\mathcal{N}=4$ supersymmetry. The W-algebra depends on the gauge algebra, and for $G=SU(2)$, it is simply given by the $\mathcal{N}=4$ super-Virasoro algebra with $c=-9$. Useful constructions of the W-algebra were given by the free field realization \cite{Bonetti:2018fqz} and by the truncation of the $W_\infty$ algebra \cite{Bonetti:2025kan}.

This SCFT/VOA correspondence raises a natural question: How does non-invertible duality symmetry act on the associated VOA? One may think that since the VOA is mostly blind to the global structure of the SCFT, the VOA is also likely the case. In this paper, however, we argue that the NIS has a subtle action on the local operator, so that it acts as an outer-automorphism on the VOA. Indeed, $\mathcal{N}=4$ super-Virasoro algebra which corresponds to the $\mathcal{N}=4$ SYM, possess an $SL(2)$ automorphism.

To demonstrate this action concretely, we consider the Schur index \cite{Gadde:2009kb,Gadde:2011uv}, which is a limit of the superconformal index \cite{Kinney:2005ej} that captures the vacuum character of the VOA. We place the NIS topological defect at a point in $S^1$ and wrapping the spatial $S^3$, which in turn is realized as the twisted trace over the Hilbert space on $S^3$. We show that this twisted index is equivalent to the vacuum character of the associated VOA twisted by the discrete action of the outer-automorphism and the additional R-symmetry. The computation of the NIS-twisted index can be done straightforwardly by using this relation via the VOA twisted character. Let us remark that the usual free field methods for computing the index is not justified since the duality defect only exists at a special strongly-coupled self-dual point. This makes it rather difficult to identify the S-duality action on the elementary fields. We leave the direct gauge theory computation from the localization as future work. Instead, we provide two independent checks of our proposal. 

The first non-trivial check of the relation comes from the large-$N$ limit. The S-duality action on the bulk fields is well-known on the gravity dual side \cite{Maldacena:1997re, Kinney:2005ej}. This allows us to easily compute the NIS-twisted Schur index, which matches the twisted character of the VOA associated with the SYM with large-$N$. 

The fusion rule gives another non-trivial check. The fusion rules of NIS defects \cite{Choi:2021kmx,Kaidi:2021xfk} are slightly extended in $\mathcal{N}=4$ SYM theory since acting the S-duality twice does not simply produce the identity but leaves the action by the center of the duality group $SL(2,\mathbb{Z})$. We argue that the latter leads to charge conjugation as well as certain discrete R-symmetry action. We confirm that the twisted character of VOA corresponding to the fusion of two defects agrees with the Schur index with the gauge charge conjugation (or outer-automorphism twisted index) \cite{Zwiebel:2011wa,Mekareeya:2012tn} and the R-symmetry twist combined. On top of these actions on the local fields, there is a global part of the fusion rule expressed as the condensation defect or the TQFT insertion \cite{Choi:2021kmx}. Since the superconformal index counts the local operators, it is blind to the global part of the fusion rule. 

The organization of this paper is as follows. In Section \ref{sec:VOA}, we define the NIS-twisted Schur index of $\mathcal{N}=4$ SYM theory with gauge group $SU(N)$ as an outer-automorphism twisted character of the W-algebra, after reviewing the NIS in $\mathcal{N}=4$ SYM theory and the associated VOA. We see that this works well in the large-$N$ by matching the VOA character against the twisted index of the supergravity. We check the consistency of our proposal by considering the fusion rules of the defect in Section \ref{sec:fusion} by matching the twisted character of the VOA and the charge conjugation index that are computed independently. In Section \ref{sec:macdonald}, we refine our computation to the Macdonald index case. In the VOA side, this is done by considering the $\mathfrak{R}$-filtration. In Section \ref{sec:conclusion}, we conclude with future perspectives. In Appendix \ref{sec:convention}, we list formulas needed to define and compute the superconformal index. In Appendix \ref{sec:U1Y}, we compute the closely related twisted index by the $U(1)_Y$ `bonus' (non-)symmetry \cite{Intriligator:1998ig}. The difference between this and the NIS twisted index is explained. In Appendix \ref{sec:U1}, the NIS twisted index of $\mathcal{N}=4$ $U(1)$ gauge theory is computed.

\section{Supersymmetric Non-invertible symmetry and outer-automorphism}
\label{sec:VOA}

\subsection{Non-invertible symmetry in $\mathcal{N}=4$ $SU(N)$ SYM}
$\mathcal{N}=4$ super Yang-Mills (SYM) theory with gauge group $SU(N)$ has the global $0$-form R-symmetry $SU(4)_R$ and the $\mathbb{Z}_N^{(1)}$ one-form symmetry. The theory enjoys Montonen-Olive $SL(2, \mathbb{Z})$ S-duality, under which the gauge coupling constant $\tau_{{\rm YM}}$ transforms as 
\begin{equation}
    \tau_{{\rm YM}} \to \frac{a \tau_{{\rm YM}} + b}{c \tau_{{\rm YM}} + d} \ ,
    \label{SL2Ztau}
\end{equation}
where $\left( \begin{smallmatrix} a & b \\ c & d \end{smallmatrix} \right) \in SL(2, \mathbb{Z})$. The fundamental elements in $SL(2, \mathbb{Z})$ are $S: \left( \begin{smallmatrix} 0 & -1 \\ 1 & 0 \end{smallmatrix} \right)$ and $T: \left( \begin{smallmatrix} 1 & 1 \\ 0 & 1 \end{smallmatrix} \right)$. The $SL(2,\mathbb{Z})$ is not a symmetry of the $SU(N)$ SYM theory even if the gauge coupling constant is set to a fixed point of the action \eqref{SL2Ztau}, since the duality action changes the global form of the gauge group. For example, the $S$ transformation takes the $SU(N)$ gauge theory to the $PSU(N)_{n=0}$ gauge theory where the subscript $n$ labels the discrete theta angle. 

The S-duality is promoted to a symmetry when combined with the gauging of the $\mathbb{Z}_N^{(1)}$ one-form symmetry and also with a possible counterterm \cite{Kaidi:2021xfk,Choi:2021kmx,Kaidi:2022uux,Choi:2022zal} that shifts the discrete theta angle. Let us denote these operations as
\begin{align}
\sigma
&: {\rm gauging~of~}\mathbb{Z}_N^{(1)}~{\rm symmetry}, \\
\tau
&: {\rm stacking ~with ~the~counterterm}~\frac{\pi}{N} \int \mathcal{P}(B),
\end{align}
where $B$ is the two-form background gauge field for $\mathbb{Z}_N^{(1)}$ and $\mathcal{P}(B)$ is the Pontryagin square of $B$. The $\sigma$ operation takes the $SU(N)$ SYM theory to the $PSU(N)_0$ SYM, while combining with the addition of the counterterm $\sigma \tau^n$ gives $PSU(N)_n$ SYM. These operations can be used to put the dual gauge group obtained by a duality transformation back to the original one. 

The simplest example is the action $\sigma S$ for $N=2$, and $\sigma^3 S$ for $N\geq 3$. The $S$ transformation acts on $SU(N)$ SYM theory with the gauge coupling $\tau_{{\rm YM}}$ as
\begin{align}
S: \quad \mathcal{T}_{SU(N)}[\tau_{{\rm YM}}, B] \rightarrow \mathcal{T}_{PSU(N)_{0}}[-1/\tau_{{\rm YM}}, B].
\end{align}
Then, for $N=2$, $\sigma$ takes this $PSU(2)_0=SO(3)_+$ SYM to $SU(2)$ with $\tau_{{\rm YM}}'=-1/\tau_{{\rm YM}}$ \cite{Aharony:2013hda}. For $N \geq 3$, $\sigma$ takes to $SU(N)$ with $\tau_{{\rm YM}}'=-1/\tau_{{\rm YM}}$, also with the background field $B$ charge-conjugated to $-B$. In order to recover the original $B$, we need to act further with $\sigma^2$, which corresponds simply to the background field charge conjugation. Thus at $\tau_{{\rm YM}} = i$, $\sigma S$ ($\sigma^3 S$) is the symmetry action of the $SU(N)$ SYM. 

As a concrete construction of the associated topological defect, we consider the half-gauging procedure \cite{Choi:2021kmx}: suppose the $SU(N)$ SYM theory with $\tau_{{\rm YM}}=i$ is on $\mathbb{R} \times M_3$, and on the half-space $\mathbb{R}_{>0} \times M_3$ we act $\sigma S$ ($\sigma^3 S$), this gives a defect at the origin of $\mathbb{R}$ occupying $M_3$. We will call this as the duality defect.

A similar construction using the $ST$ transformation leads to the triality defect of the theory with the gauge coupling $\tau_{{\rm YM}}=e^{2\pi i /3}$. The $ST$ transformation gives
\begin{align}
ST :
\quad \mathcal{T}_{SU(N)}[\tau_{{\rm YM}}, B] \rightarrow \tau(\mathcal{T}_{PSU(N)_{0}}[-1/(\tau_{{\rm YM}}+1), B]).
\end{align}
Then the operations $\sigma \tau^{-1}$ for $N=2$ and $\sigma^3 \tau^{-1}$ for $N\geq 3$ set the theory back to the original theory with $\tau_{{\rm YM}}'= -\frac{1}{\tau_{{\rm YM}}+1}$. The fixed point of the gauge coupling constant is $\tau_{{\rm YM}} = e^{2\pi i /3}$. By similar half-gauging, the associated triality defect is constructed. See \cite{Kaidi:2022uux} for a complete list of topological defects.

These defects in general break the supersymmetry: under the $S$ and $ST$ transformations, the supercharges $\mathcal{Q}^I$ get a nontrivial phase \cite{Kapustin:2006pk}
\begin{align}
 S ~(\textrm{or } ST): \quad \mathcal{Q}^I \to \omega^{-\frac{1}{2}} \mathcal{Q}^I,
 \label{KW}
\end{align}
where $\omega= e^{2\pi i/k}$ with $k=4$ for $S$ and $k=6$ for $ST$, while $\tau, \sigma$ do not act on local operators. To consider the superconformal index with such defects inserted, we need to preserve some amount of supersymmetry. Here we choose to combine the duality transformation with a discrete action in $SU(4)_R$. Denoting the $SU(4)_R$ Cartan charges as $R_1$, $R_2$, and $R_3$ (See Appendix \ref{sec:convention} for conventions), we consider the following one-parameter family
  \begin{align}
  R_\zeta
   =     \frac{R_1}{2}+ R_2 + (2\zeta +1)\frac{R_3}{2} \ , 
  \end{align}
where $\zeta \in \mathbb{R}$. Indeed, under the action by $\mathcal{R}_\zeta$ with  $\omega^{R_\zeta}=e^{\frac{2 \pi i }{k}R_\zeta}$, the supercharges transform as
\begin{align}
   \mathcal{R}_\zeta: \quad  (\mathcal{Q}^1, \mathcal{Q}^2, \mathcal{Q}^3, \mathcal{Q}^4) \rightarrow (\omega^{\frac{1}{2}} \mathcal{Q}^1, \omega^{\frac{1}{2}}\mathcal{Q}^2, \omega^{\zeta-\frac{1}{2}} \mathcal{Q}^3, \omega^{- \zeta -\frac{1}{2}} \mathcal{Q}^4) \ .
    \label{SU(4)R}
\end{align}
Let us denote the (invertible) topological operator that acts as above as $\mathcal{R}_\zeta$.  
By combining with \eqref{KW}, this transformation preserves at least $\mathcal{N}=2$ supersymmetry among $\mathcal{N}=4$:
\begin{align}
    (S \textrm{ or } ST) \cdot \mathcal{R}_\zeta \, : \quad (\mathcal{Q}^1, \mathcal{Q}^2, \mathcal{Q}^3, \mathcal{Q}^4) \rightarrow (\mathcal{Q}^1, \mathcal{Q}^2, \omega^{\zeta - 1} \mathcal{Q}^3, \omega^{-\zeta-1} \mathcal{Q}^4) \ .
    \label{r_k}
\end{align}
Note that when $\zeta=0$, $R_{\zeta=0}$ is the $U(1)_r$ of $\mathcal{N}=2$ R-symmetry. Also note that when $\zeta=1$, this action preserves $\mathcal{N}=3$ supersymmetry\footnote{This action is considered in \cite{Nishinaka:2016hbw,Bourton:2018jwb,Kaidi:2022uux}.} .

In the following, we will refer to the supersymmetry-preserving duality defect and triality defect as $\mathcal{D}_d$ and $\mathcal{D}_t$, respectively
\begin{align}
\mathcal{D}_d
: \sigma^n S \, \mathcal{R}_{\zeta} , ~~~~~~~
\mathcal{D}_t
: \sigma^n \tau^{-1} ST \, \mathcal{R}_{\zeta},
\end{align}
where $n=1$ for $N=2$ and $n=3$ for $N\geq 3$. As we will see later and as in \cite{Kaidi:2022uux}, these satisfy non-invertible fusion rules. 

\subsection{Non-invertible symmetry-twisted Schur index}
\label{sec:index}
Let us consider the NIS-twisted superconformal index of $\mathcal{N}=4$ SYM theory, which is our main object in the paper, as that of an $\mathcal{N}=2$ SCFT. Since the supercharges $(\mathcal{Q}_1, \mathcal{Q}_2)$ are not broken by the defects ($\mathcal{D}_{d, t}$), we can use them to define the $\mathcal{N}=2$ index \cite{Gadde:2009kb,Gadde:2011uv}. As an $\mathcal{N}=2$ SCFT, $U(1)_r \times SU(2)_R$ ($\subset SU(4)_R$) is the R-symmetry and its commutant $SU(2)_a$ can be viewed as a flavor symmetry. 

We focus on the Schur limit of the superconformal index. The Schur index is defined as the trace over the Hilbert space on $S^3$
\begin{equation}
\mathcal{I}(q,a)
 =      {\rm Tr}_{\mathcal{H}(S^3)} (-1)^F q^{E-R} a^f,
\end{equation}
where $a$ is the fugacity of $SU(2)_a$, $E$ is the conformal dimension, and $R$ is the generator of $SU(2)_R$ given by $R=R_1/2$ in terms of $SU(4)_R$ charges. The Schur index gets contributions from the Schur operators that satisfy $\delta_1 = \delta_2=0$, where 
  \begin{align}
  \begin{split}
  \delta_1
   &=     \frac{1}{2} \{ \mathcal{Q}^1_-, \mathcal{S}_1^- \}
   =     E - 2j_1 -2 R - r, \\
   \delta_2
   &=     \frac{1}{2} \{ \tilde{\mathcal{Q}}_{2 \dot{-}}, \tilde{\mathcal{S}}^{2 \dot{-}} \}
   =     E - 2j_2 -2 R + r.
   \label{delta12}
   \end{split}
  \end{align}
For the free $\mathcal{N}=4$ vector multiplet, the elementary fields which satisfy the Schur conditions are summarized in Table \ref{tab:chargesN=4} with their charge assignments. From this, we read off the single-letter Schur index given as 
\begin{align}
    i(q, a)
     = \frac{- 2q + q^{\frac{1}{2}}(a+\frac{1}{a}) }{1-q}.
    \label{singleSchur}
\end{align}
Then the index for the SYM theory is computed by taking the plethystic exponent (PE) and integrating over the gauge group as 
\begin{equation}
\mathcal{I}(q, a)
 =     \int [d U] {\rm PE}[i(q, a)\chi_{{\rm adj}}(u)],
 \label{SchurNIS}
\end{equation}
where $U$ denotes an element of the gauge group $G$ and $[d U]$ is the invariant Haar measure. Also, $\chi_{{\rm adj}}(u)$ denotes the character of the adjoint representation of $G$. 

\begin{table}
    \centering
    \begin{tabular}{|c||c|c|c|c|c|c|c|c|c||c|c|}
         \hline
         & $E$ & $j_1$ & $j_2$ & $R_1$ & $R_2$  & $R_3$ & $R_\zeta$ & $Y$ & $r_0$ & $i$ & $i_{\tilde{R}_0}$  \\
         \hline \hline
     $Y$ & $1$ & $0$ & $0$ & $1$  & $-1$  & $1$ & $\zeta$ & $0$ & $-1$ & $q^{\frac{1}{2}} a$ & $q^{\frac{1}{2}} a \eta^{-1+\zeta}$\\
          \hline
     $Z$ & $1$ & $0$ & $0$ & $1$  & $0$  & $-1$ & $-\zeta$ & $0$ & $-1$ & $q^{\frac{1}{2}} a^{-1}$ & $q^{\frac{1}{2}} a^{-1} \eta^{-1-\zeta}$ \\
          \hline
     $\bar{\lambda}^1_{\dot{+}}$ & $\frac{3}{2}$ & $0$ & $\frac{1}{2}$ & $1$  & $0$  & $0$ & $\frac{1}{2}$ & $1$ & $-\frac{1}{2}$ & $-q$ & $-q$ \\
          \hline 
     $\lambda_{2+}$ & $\frac{3}{2}$ & $\frac{1}{2}$ & $0$ & $1$  & $-1$  & $0$ & $-\frac{1}{2}$ & $-1$ & $\frac{1}{2}$ & $-q$ & $-q$ \\
          \hline
    \end{tabular}
    \caption{Charges of the Schur letters in $\mathcal{N}=4$ vector multiplet under the R-symmetry and the $U(1)_Y$ action. $(Y,Z)$ are complex scalars, and $\lambda_I$ are Weyl fermions in $\mathcal{N}=4$ vector multiplet. $r_0$ is the center of the R-symmetry, and $R_\zeta = \frac{R_1}{2}+R_2+(2\zeta+1)\frac{R_3}{2}$. See Appendix \ref{sec:convention} for notations. $i_{\tilde{R}_0}$ is the contribution to the single-letter index of $\tilde{R}_0$ twist.}
    \label{tab:chargesN=4}
\end{table}

We consider the insertion of the defect $\mathcal{D}_{d,t}$ on a point in $S^1$ and wrapping the spatial $S^3$. The index is given by the twisted trace in the same Hilbert space. This detects only the local gauge-invariant operators. Thus the global part of $\mathcal{D}_{d,t}$, $\sigma$, $\tau$ and $T$, does not contribute. Therefore, the defect Schur index is given by:
\begin{equation}
\mathcal{I}_{\mathcal{D}_{d,t}}(q, a; \omega; \zeta)
 =     {\rm Tr}_{\mathcal{H}_S(S^3)} (-1)^F q^{E-R} a^f \omega^{s+R_\zeta},
 \label{NISSchurindex}
\end{equation}
where $s$ is the charge under the $S$ transformation such that the charge of $\mathcal{Q}^I$ is normalized $-\frac{1}{2}$, and $\omega= e^{2\pi i/k}$ with $k=4$ for $\mathcal{D}_d$ and $k=6$ for $\mathcal{D}_t$.

A subtlety in computing the twisted index is that the $S$ transformation exchanges electric and magnetic variables, which makes it difficult to trace how single-letter transforms under $S$ in a non-Abelian gauge theory. Instead, we will argue in Section \ref{sec:VOAauto} that the automorphism-twisted character of VOA computes this index properly. We remark that there is no such subtlety in the case of free $U(1)$ vector multiplet. As a warm-up example, we compute the defect Schur index of free $\mathcal{N}=4$ gauge theory in Appendix \ref{sec:U1}. 

\paragraph{$U(1)_Y$ symmetry}
Related to the aforementioned subtlety of the $S$ transformation, there is another action on $\mathcal{N}=4$ gauge theory, which is the so-called $U(1)_Y$ bonus symmetry \cite{Intriligator:1998ig}. This `symmetry' is a remnant of $SL(2, \mathbb{R})$ symmetry of the Type IIB supergravity, which holds classically but broken quantum mechanically. The $U(1)_Y$ bonus symmetry is defined by the action on the elementary fields as the following phase rotation:
\begin{align}
F^\pm 
\rightarrow e^{\mp 2 i\phi}  F^\pm, ~~~~
\lambda_I 
\rightarrow e^{- i \phi}\lambda_I, ~~~~
(X,Y,Z)
\rightarrow (X,Y,Z), 
\label{U1Y}
\end{align}
while on the supercharges as 
\begin{align}
\mathcal{Q}^I &\rightarrow e^{-i \phi}\mathcal{Q}^I.
\end{align}
It is obvious that for non-Abelian gauge theory, the $U(1)_Y$ action is not a symmetry in the sense that it does not preserve the equations of motion. For example, the on-shell supersymmetry transformation of a Weyl fermion gives the field strength and commutator of the scalars; these terms have different charge under $U(1)_Y$. Note that only when $\phi=\pi$, the equations of motion are preserved. Thus, this $\mathbb{Z}_2 \subset U(1)_Y$ is an actual symmetry. For Abelian gauge theory, the whole $U(1)_Y$ is indeed a symmetry of the equations of motion. 

The action on the supercharges is the same as that of the $S$ and $ST$ transformations when $\phi=\pi/k$ with $k=4$ for $S$ and $k=6$ for $ST$. 
Thus, one may be tempted to consider the twisted index by $U(1)_Y$ with angle $\pi/k$ (and with $R_\zeta$) to see the relevance with the non-invertible twisted index \eqref{NISSchurindex}:
\begin{equation}
\mathcal{I}_{Y}(q, a; \omega; \zeta)
 =     {\rm Tr}_{\mathcal{H}_S(S^3)} (-1)^F q^{E-R} a^f \omega^{\frac{Y}{2}+R_\zeta},
 \label{YSchurindex}
\end{equation}
where again $\omega=e^{2\pi i/k}$ and $Y$ is normalized such that the charge of $F^+$ is $-1$. (See Table \ref{tab:chargesN=4}.) Since the $U(1)_Y$ action is not a symmetry of $\mathcal{N}=4$ $SU(N)$ SYM, this twisted trace is ill-defined. We will not consider this here and refer to Appendix \ref{sec:U1Y} for a ``formal" computation of this object. However, $U(1)_Y$ provides a hint for finding the right action on VOA. 

\subsection{VOA automorphism}
\label{sec:VOAauto}
We consider a W-algebra $\mathcal{W}_{\infty}^{s,s}$ to construct the VOA associated with $\mathcal{N}=4$ $SU(N)$ SYM \cite{Bonetti:2025kan, Gaberdiel:2025eaf}. It is a non-linear W-algebra, containing the small $\mathcal{N}=4$ super Virasoro algebra as a subalgebra. Therefore the operators in $\mathcal{W}_{\infty}^{s,s}$ form multiplets under $\mathfrak{psl}(2|2)$, which is the finite subalgera of the $\mathcal{N}=4$ super-Virasoro. The $\mathfrak{psl}(2|2)$ algebra has a bosonic subalgebra $\mathfrak{sl}(2)_{z}\oplus\mathfrak{sl}(2)_{y}$, where $\mathfrak{sl}(2)_{z}$ corresponds to the (global part of the) $2d$ conformal symmetry and $\mathfrak{sl}(2)_{y}$ being the R-symmetry. We label each operator by the conformal weight $h$ and $\mathfrak{sl}(2)_{y}$ spin $j$. All the operators of $\mathcal{W}_{\infty}^{s,s}$ are written as combinations of normal-ordered products of the strong generators listed in Table \ref{tab:Wgenerator}.
\begin{table}[h]
    \centering
    \begin{tabular}{|c||c|c|c|c|c|c|c|c|}
        \hline
        $\mathfrak{psl}(2|2)$ multiplet & \multicolumn{4}{|c|}{$\mathbb{J}$} & \multicolumn{4}{|c|}{$\mathbb{W}_{p>2}$} \\\hline
        Operators & $J$ & $G$ & $\tilde{G}$ & $T$ & $W_{p}$ & $G_{W_{p}}$ & $\tilde{G}_{W_{p}}$ & $T_{W_{p}}$ \\\hline
        Conformal weight($h$) & $1$ & $\frac{3}{2}$ & $\frac{3}{2}$ & $2$ & $\frac{p}{2}$ & $\frac{p+1}{2}$ & $\frac{p+1}{2}$ & $\frac{p+2}{2}$ \\\hline
        $\mathfrak{sl}(2)_{y}$ spin($j$) & $1$ & $\frac{1}{2}$ & $\frac{1}{2}$ & $0$ & $\frac{p}{2}$ & $\frac{p-1}{2}$ & $\frac{p-1}{2}$ & $\frac{p-2}{2}$\\\hline
    \end{tabular}
    \caption{Strong generators of $\mathcal{W}_{\infty}^{s,s}$ and their charges.}
    \label{tab:Wgenerator}
\end{table}
The $\mathfrak{psl}(2|2)$ multiplet $\mathbb{J}$ consists of strong generators of $\mathcal{N}=4$ super Virasoro algebra. There is an additional set of strong generators denoted as $\mathbb{W}_{p}$ for each $p=3,4,\cdots$. The $\mathfrak{psl}(2|2)$ primary $W_{p}$ is a Grassmann even Virasoro primary.

The universal $\mathcal{W}_{\infty}^{s,s}$ algebra has central charge $c$ as the only free parameter. It is conjectured that when the central charge is set to be
\begin{align}
    c=3(1-N^{2}), ~~~ N=2,3,\cdots,
\end{align}
the $\mathcal{W}_{\infty}^{s,s}$ algebra contains null states. In particular, all generators in the multiplet $\mathbb{W}_{p}$ for $p>N$ are null. Let $\mathcal{I}_{N}$ denote an ideal of null states generated by $W_{p>N}$. The simple quotient of $\mathcal{W}_{\infty}^{s,s}$ by $\mathcal{I}_{N}$ is conjectured to be isomorphic to the VOA associated to $\mathcal{N}=4$ $SU(N)$ SYM. We denote this VOA as $\mathcal{V}(A_{N-1})$.

One crucial property for us is that the algebra $\mathcal{W}_{\infty}^{s,s}$ enjoys an outer-automorphism $\mathfrak{sl}(2)_{\rm{out}}$ under which $G$($G_{W_{p}}$) and $\tilde{G}$($\tilde{G}_{W_{p}}$) transform as a doublet. We denote charges under the Cartan $\mathfrak{gl}(1)_{r}$ of $\mathfrak{sl}(2)_{\rm out}$ as $r$, so that $G$($G_{W_{p}}$) and $\tilde{G}$($\tilde{G}_{W_{p}}$) has charge $r$ as $+\frac12$ and $-\frac12$ respectively. This outer-automorphism is inherited by $\mathcal{V}(A_{N-1})$ too, which allows us to define a character of $\mathcal{V}(A_{N-1})$ twisted by this automorphism. In particular, we consider the following super-character twisted by the automorphism as well as $\mathfrak{sl}(2)_y$ with the weight $\omega^\zeta$,
\begin{align}
    \chi_{\mathcal{V}(A_{N-1})}(q,a;\omega;\zeta)={\rm Tr}(-1)^{F}q^{h}a^{j}\omega^{2r+\zeta j},
    \label{VOAtwcharacter}
\end{align}
where $F$ is the fermion number.\footnote{We often do not mention the label $F$ below, because $F$ is related to $r$ as $F=2r$ for $\mathcal{W}_{\infty}^{s,s}$ and $\mathcal{V}(A_{N-1})$.} This twisted supercharacter can be seen as the automorphism twist with the fugacity shift as $a \rightarrow a \omega^{\zeta}$.

This twist action by $\omega^{2r+\zeta j}$ with $\omega=e^{2 \pi i/k}$ is actually the $U(1)_Y$ with $\phi=\pi/k$ plus $R_\zeta$ action, which we denote as $\mathcal{Y}= \omega^{\frac{Y}{2}+R_\zeta}$, in the $4d$ gauge theory that carries over to VOA. To see this, consider the BRST construction of VOA in \cite{Beem:2013sza} where the gauge-invariant operators constructed from the Schur letters are identified with the generators of VOA. For example, the following operators in the $4d$ theory are mapped to the VOA generators as
\begin{align}
\begin{split}
({\rm Tr}(Y^2), {\rm Tr}(YZ), {\rm Tr}(Z^2)) 
&\leftrightarrow (J_+, J_0, J_-), \\
({\rm Tr}(\bar{\lambda}^1Y), {\rm Tr}(\bar{\lambda}^1Z), {\rm Tr}(\lambda_2Y), {\rm Tr}(\lambda_2Z)) 
&\leftrightarrow (G_1, G_2, \tilde{G}_1,\tilde{G}_2).
\end{split}
\end{align}
The $U(1)_Y$ with $\phi=\pi/k$ rotation combined with $\omega^{R_\zeta}$ on the Schur letters translates to the action
\begin{align}
\begin{split}
(J_+, J_0, J_-)
&\mapsto (\omega^{2\zeta} J_+, J_0, \omega^{-2\zeta} J_-), \\
(G_1, G_2, \tilde{G}_1, \tilde{G}_2)
&\mapsto (\omega^{1+\zeta} G_1, \omega^{1-\zeta}G_2, \omega^{-1+\zeta}\tilde{G}_1, \omega^{-1-\zeta}\tilde{G}_2).
\end{split}
\end{align}
This is the $\mathfrak{sl}(2)_r$ and $\mathfrak{sl}(2)_y$ action $\omega^{2r + \zeta j}$. Note that after passing to VOA, the action $\mathcal{Y}$ survives as a non-trivial automorphism. This allows us to define the twisted index, even though $U(1)_Y$ is not a symmetry of the full 4d theory.

Now, we argue that what we are computing in the VOA coincides with the NIS twisted index \eqref{NISSchurindex}, rather than the $U(1)_Y$ twisted index \eqref{YSchurindex}. Let us discuss the $\mathcal{D}_d$ and $\mathcal{Y}$ actions with $\omega=e^{2 \pi i /k}$, $k=4$, and omit their global part. The comparison between $\mathcal{D}_t$ and $\mathcal{Y}$ with $k=6$ goes as well. The point is that $\mathcal{D}_d$ and $\mathcal{Y}$ act on all the supercharges in the same way as a phase rotation. This means that $\mathcal{D}_d$ and $\mathcal{Y}$ act in the same way on the other generators of $\mathfrak{psu}(2,2|4)$. Then, define $A=\mathcal{Y}^{-1}\mathcal{D}_d$, the difference of $\mathcal{D}_d$ and $\mathcal{Y}$. This satisfies $[A, X]=0$ for any $X \in \mathfrak{psu}(2,2|4)$. By Schur's lemma, this leads to that for any superconformal primary $\mathcal{O}$ in irreducible representation of $\mathfrak{psu}(2,2|4)$, $A \mathcal{O} = c \, \mathcal{O}$ where $c$ is at least a phase. Any superconformal descendant $\mathcal{O}_d$ from $\mathcal{O}$ has the same $c$: $A \mathcal{O}_d = c \, \mathcal{O}_d$. This argument shows that for any multiplet, $A$ acts as a phase $c$. 

A special case where we can determine $c$ is the $\mathcal{N}=4$ stress-tensor multiplet. The stress-tensor is preserved both by $\mathcal{D}_d$ and $\mathcal{Y}$. Thus, for the stress-tensor multiplet, $c=1$. 

For the $SU(2)$ case, all the strong generators in $\mathcal{V}(A_1)$ come from $4d$ stress-tensor multiplet. Thus, $\mathcal{D}_d$ and $\mathcal{Y}$ acts in the same way on the VOA generators. This proves that the twisted character \eqref{VOAtwcharacter} computes the NIS twisted index \eqref{NISSchurindex} for $SU(2)$.

For $SU(N)$ with $N \geq 3$, the multiplet $\mathbb{W}_p$ in $\mathcal{V}(A_{N-1})$ comes from $4d$ other multiplet, say $w_p$, where $c_p$ ($A w_p = c_p w_p$) is not fixed for all $p$. The large $N$ consideration almost answers this. For large $N$, the algebra is $\mathcal{W}_\infty^{s,s}$ and all $\mathbb{W}_p$ multiplets are not truncated. From the gravity dual which we will discuss in Section \ref{sec:largeN} in detail, we know the action $S$ or $SL(2,\mathbb{Z})$ is lifted to $SL(2, \mathbb{R})$, whose Cartan is $U(1)_Y$. Thus, $c_p=1$ for all $p$. We conjecture this is preserved for finite $N$.

The calculation of the supercharacter can be done straightforwardly by counting the $\mathfrak{psl}(2|2)$ primaries. We have two types of $\mathfrak{psl}(2|2)$ multiplets: short and long multiplets. Consider a $\mathfrak{psl}(2|2)$ primary $X$ with conformal weight $h_{X}$, $\mathfrak{sl}(2)_{y}$ spin $j_{X}$ and $\mathfrak{gl}(1)_{r}$ charge $r_{X}$. When $h_{X}=j_{X}$, the action of fermionic generator of super Virasoro algebra is terminated, resulting a short multiplet. On the other hand, when $h_{X}>j_{X}$, the resulting multiplet is not shortened. However, when $j_{X}=0$ or $j_{X}=\frac{1}{2}$, the long multiplet contains negative spin or degenerate operators, so they should be considered more carefully. All of the above discussion is summarized into the following contributions of short and long multiplets,
\begin{align}
    \mathfrak{S}_{j}(q,a,\omega)
    &= \frac{1}{1-q}{\bigg [}q^{j}\chi_{j}(a)-q^{j+\frac12}\chi_{j-\frac12}(a){\big(}\omega+\frac1\omega{\big)}+q^{j+1}\chi_{j-1}(a){\bigg ]},\label{multiplets}\\
    \mathfrak{L}_{h,j>0}(q,a,\omega)
    &= \frac{1}{1-q}{\bigg [}q^{h}\chi_{j}(a)-q^{h+\frac12}\chi_{j-\frac12}(a){\big(}\omega+\frac1\omega{\big)}-q^{h+\frac12}\chi_{j+\frac12}(a){\big(}\omega+\frac1\omega{\big)}\nonumber\\
    &+q^{h+1}\chi_{j}(a){\big(}\omega^{2}+2+\frac{1}{\omega^{2}}{\big)}+q^{h+1}\chi_{j-1}(a)+q^{h+1}\chi_{j+1}(a)\nonumber\\
    &-q^{h+\frac32}\chi_{j-\frac12}(a){\big(}\omega+\frac1\omega{\big)}-q^{h+\frac32}\chi_{j+\frac12}(a){\big(}\omega+\frac1\omega{\big)}+q^{h+2}\chi_{j}(a){\bigg ]},\label{multipletl}\\
    \mathfrak{L}_{h,0}(q,a,\omega)
    &= \frac{1}{1-q}{\bigg [}q^{h}-q^{h+\frac12}\chi_{\frac12}(a){\big(}\omega+\frac1\omega{\big)}+q^{h+1}{\big(}\omega^{2}+1+\frac{1}{\omega^{2}}{\big)}\nonumber\\
    &+q^{h+1}\chi_{1}(a)-q^{h+\frac32}\chi_{\frac12}(a){\big(}\omega+\frac1\omega{\big)}+q^{h+2}{\bigg ]}.
    \label{multipletl0}
\end{align}
Here, $\chi_{j}(a)$ is spin-$j$ $\mathfrak{sl}(2)$ character,
\begin{align}
    \chi_{j}(a)=a^{-2j}+\cdots+a^{2j},
\end{align}
for $j\geq0$ and $\chi_{j<0}(a)=0$. The supercharacter \eqref{VOAtwcharacter} can be expanded into contributions of long and short multiplets
\begin{align}
    \chi_{\mathcal{V}(A_{N-1})}(q,a;\omega;\zeta)&=\sum_{j,r}n_{j,j,r}(-1)^{F}\omega^{2r}\mathfrak{S}_{j}(q,a\omega^\zeta,\omega)+\sum_{h,j,r}n_{h,j,r}(-1)^{F}\omega^{2r}\mathfrak{L}_{h,j}(q,a\omega^\zeta,\omega),\label{VOAtwcharacterexp}
\end{align}
where $n_{h,j,r}$ is the number of $\mathfrak{psl}(2|2)$ primaries in $\mathcal{V}(A_{N-1})$ with quantum numbers $(h,j,r)$. For later use, we list the low-lying spectrum of $\mathfrak{psl}(2|2)$ primaries up to $h=4$ in Table \ref{tab:pslprimaries}. (See Table 12 of \cite{Bonetti:2025kan} for more complete expressions of the primaries.)

\begin{table}[t]
    \centering
    \begin{tabular}{|c||c||c||c|c||c|c|c||c|c|c||c|c|c|c||c|c|c|c|c|}
    \hline
      $h$ & $1$ & $\frac32$ & \multicolumn{2}{|c||}{$2$} & \multicolumn{3}{|c||}{$\frac52$} & \multicolumn{3}{|c||}{$3$} & \multicolumn{4}{|c||}{$\frac72$} & \multicolumn{5}{|c|}{$4$}\\\hline
      $j$ & $1$ & $\frac32$ & $2$ & $0$ & $\frac52$ & $\frac32$ & $\frac12$ & $3$ & $2$ & $1$ & $\frac72$ & $\frac52$ & $\frac32$ & $\frac12$ & $4$ & $3$ & $2$ & $1$ & $0$\\\hline\hline
      $N=2$ & 1 & 0 & 1 & 0 & 0 & 0 & 0 & 1 & 0 & 1 & 0 & 0 & 0 & 0 & 1 & 0 & 1 & 0 & 1 \\\hline
      $N=3$ & 1 & 1 & 1 & 1 & 1 & 1 & 0 & 2 & 0 & 2 & 1 & 1 & 2 & 1 & 2 & 1 & 3 & 1 & 3 \\\hline
      $N=4$ & 1 & 1 & 2 & 1 & 1 & 1 & 1 & 3 & 1 & 3 & 2 & 2 & 3 & 2 & 4 & 2 & 7 & {\tiny $3-\omega-\frac1\omega$} & 5 \\\hline\hline
      $\mathcal{W}_{\infty}^{s,s}$ & 1 & 1 & 2 & 1 & 2 & 1 & 1 & 4 & 1 & 4 & 4 & 3 & 5 & 3 & 7 & 4 & 11 & {\tiny $5-\omega-\frac1\omega$} & 7 \\\hline
    \end{tabular}
    \caption{Spectra of $\mathfrak{psl}(2|2)$ primaries of $\mathcal{V}(A_{N-1})$, when $N=2, 3, 4$ and $\mathcal{W}_{\infty}^{s,s}$ without truncation. Each entry is $\sum_{r}n_{h,j,r}(-1)^{F}\omega^{2r}$ in \eqref{VOAtwcharacterexp} with fixed $h$ and $j$. For comparison, we show the primary spectrum for $\mathcal{W}_{\infty}^{s,s}$. It is clear that some states in $\mathcal{V}(A_{N-1})$ become null after the simple quotient.}
    \label{tab:pslprimaries}
\end{table}

\paragraph{$\mathcal{V}(A_1)$ case}
Let us consider the $\mathcal{V}(A_1)$ algebra. In this case, the strong generators are those in $\mathbb{J}$. Under the outer-automorphism $\mathfrak{sl}(2)_{{\rm out}}$, $G$ and $\tilde{G}$ have charge $r=\frac{1}{2}$ and $-\frac{1}{2}$.

From the Table \ref{tab:pslprimaries} and \eqref{multiplets}-\eqref{multipletl0}, it is straightforward to compute the twisted character:
\begin{align}
\chi_{\mathcal{V}(A_1)}(q,a;\omega;\zeta)
&=    1+\mathfrak{S}_{1}+\mathfrak{S}_{2}+\mathfrak{S}_{3}+\mathfrak{L}_{3,1}+\mathfrak{S}_{4}+\mathfrak{L}_{4,2}+\mathfrak{L}_{4,0}+\cdots \nonumber \\
&=    1 + \chi_1 q - (\omega + \omega^{-1}) \chi_{\frac{1}{2}} q^{\frac{3}{2}} + (\chi_2 + \chi_1 + 1)q^2  \nonumber \\
& ~~~- (\omega + \omega^{-1})(\chi_{\frac{3}{2}}+\chi_{\frac{1}{2}})q^{\frac{5}{2}} + (\chi_3+\chi_2+3\chi_1+1)q^3 \nonumber \\
&~~~~ -(\omega+\omega^{-1})(\chi_{\frac52}+2\chi_{\frac32}+2\chi_{\frac12}) q^{\frac72} \nonumber \\
&~~~~ +(\chi_{4}+\chi_{3}+4\chi_{2} +(\omega^{2}+5+\omega^{-2})\chi_{1}+3)q^4 + \ldots,
\label{chiA1}
\end{align}
where for the $\mathfrak{sl}(2)$ character we have omitted the argument, which in this twisted case is $\chi_j(a \omega^\zeta)$. Also, the long and short multiplet contributions we have omitted too the argument $\mathfrak{S}_j(q, a \omega^{\zeta}; \omega)$, and $\mathfrak{L}_{h,j}(q, a \omega^{\zeta}; \omega)$.

When $\omega = e^{\frac{2 \pi i}{4}}= i$, which corresponds to the duality defect, we conjecture the following closed form of the twisted character for arbitrary $\zeta$:
\begin{align}
    \chi_{\mathcal{V}(A_1)}(q, a; i; \zeta)
    =     \prod_{k=1}^\infty \frac{(1 - q^{4 k}) (1 - (-1)^\zeta q^{4 k} a^2) (1 - (-1)^{-\zeta}q^{4 k} a^{-2})}{(1 - q^k)(1 -(-1)^\zeta q^k a^2)(1 -(-1)^{-\zeta} q^k a^{-2})}
\end{align}

It is interesting to see that this twisted character has a nice modular property, similar to the case the usual untwisted Schur index \cite{Beem:2017ooy, Kang:2021lic, Pan:2021mrw, Beem:2021zvt,Zheng:2022zkm}. Let us focus on the case with $\zeta=1$, corresponding to $\mathcal{N}=3$ preserving defect, and $\omega=i$ and  $a=1$. By multiplying the overall factor $q^{\frac{3}{8}}$ the twisted character $\hat{\chi}_{\mathcal{V}(A_1)}$
\begin{align}
    \hat{\chi}_{\mathcal{V}(A_1)}(q, 1; i; \zeta=1)
     =     q^{\frac{3}{8}} \Big( 1 - q + q^2 - 2 q^3 + 4 q^4 - 5 q^5 + \dots \Big)
\end{align}
satisfies the modular differential equation:
\begin{align}
    \left( D^{(2)}_q + \frac{1}{24} \tilde{\Theta}_{(0,1)} D^{(1)}_q - \frac{3}{128} \tilde{\Theta}_{(1,1)} \right) \hat{\chi}_{\mathcal{V}(A_1)}(q, 1; i; 1)
     =     0,
\end{align}
where 
\begin{align}
    D^{(1)}_q 
     =     q \partial_q, ~~~
    D^{(2)}_q
     =     (q \partial_q + 2 \mathbb{E}_2)q \partial_q
\end{align}
and $\mathbb{E}_2$ is the second Eisenstein series.
Also the theta functions are given by
\begin{align}
    \tilde{\Theta}_{(r,s)}
     =     (-1)^{r+s} (\theta_4^{4r} \theta_3^{4s}+ \theta_3^{4r} \theta_4^{4s}),
\end{align}
where 
\begin{align}
\theta_{3} = \sum_{n=-\infty}^\infty q^{\frac{n^2}{2}}, ~~~
\theta_4 = \sum_{n=-\infty}^\infty (-1)^n q^{\frac{n^2}{2}}.
\end{align}

\subsection{Large-$N$ limit}
\label{sec:largeN}
Let us consider the large-$N$ limit of $\chi_{\mathcal{V}(A_{N-1})}$ as a consistency check. Since $g_{{\rm YM}}$ is fixed as a non-zero finite value, taking the large-$N$ limit places the theory in the regime of AdS/CFT \cite{Maldacena:1997re}. In AdS/CFT duality, the dual description of $\mathcal{N}=4$ $SU(N)$ SYM in the large-$N$ limit is a weakly coupled type IIB supergravity theory compactified on $AdS_{5}\times S^{5}$. The states of the bulk dual theory are labeled by the same set of quantum numbers as those of $\mathcal{N}=4$ SYM. In the bulk side, the superconformal multiplet $S_{n}$ is constructed from lowest-weight state with $E=n$, $j_{1}=j_{2}=0$ and having Dynkin label $(0,n,0)$ under $SU(4)_{R}$, for $n\geq2$. The multiplet structure can also be found from direct KK reduction on $S^{5}$ \cite{Kinney:2005ej, Gunaydin:1984fk,Bonetti:2016nma}. In either case, the Schur multiplet consists of states listed in Table \ref{tab:kkschur}. In type IIB supergravity, there is an R-symmetry $U(1)_{Y}$ rotating chiral supercharges in a doublet \cite{Intriligator:1998ig}. We fix their charges as $\pm\frac12$. We keep track of that charge denoted as $s$, since it is the charge under the $S$ transformation in the dual $\mathcal{N}=4$ SYM.
\begin{table}
    \centering
    \begin{tabular}{|c|c|c|c|c|c||c|}
        \hline
        $E$ & $j_{1}$ & $j_{2}$ & $R$ & $r$ & $f$ & $s$  \\\hline\hline
        $n$ & $0$ & $0$ & $\frac n2$ & 0 & $\frac n2$ & 0\\\hline
        $n+\frac12$ & $0$ & $\frac12$ & $\frac n2$ & $\frac12$ & $\frac n2-\frac12$ & $\frac12$\\\hline
        $n+\frac12$ & $\frac12$ & $0$ & $\frac n2$ & $-\frac12$ & $\frac n2-\frac12$ & $-\frac12$\\\hline
        $n+1$ & $\frac12$ & $\frac12$ & $\frac n2$ & 0 & $\frac n2-1$ & 0\\\hline
    \end{tabular}
    \caption{Components of supersymmetric multiplet constructed from the lowest-weight state $S_{n\geq2}$, satisfying the Schur condition.}
    \label{tab:kkschur}
\end{table}

Let us calculate the defect Schur index \eqref{NISSchurindex} on the supergravity side. Here, for the strict large $N$ limit, the full multi-trace index can be computed by taking the Plethystic exponential of the single-trace index. 
The single-trace (or single graviton) Schur index is calculated as sum of contributions from $S_{n\geq2}$ from the KK reduction,
\begin{align}
    i(q,a;\omega;\zeta)=\sum_{n=2}^{\infty}\frac{q^{\frac n2}\chi_{\frac n2}(a\omega^{\zeta})-q^{\frac n2+\frac12}(\omega+\omega^{-1})\chi_{\frac n2-\frac12}(a\omega^{\zeta})+q^{\frac n2+1}\chi_{\frac n2-1}(a\omega^{\zeta})}{1-q}.
\end{align}
The full Schur index is now given by the PE of the above `single graviton' index. It can be written as an infinite product,
\begin{align}
    \mathcal{I}_{\mathcal{D}_{d,t}}(q,a;\omega;\zeta)=\prod_{n=2}^{\infty}\frac{{\displaystyle\prod_{m=-\frac n2+\frac12}^{\frac n2-\frac12}(q^{\frac n2+\frac12}a^{2m}\omega^{1+2\zeta m};q)(q^{\frac n2+\frac12}a^{2m}\omega^{-1+2\zeta m};q)}}{{\displaystyle\prod_{m=-\frac n2}^{\frac n2}}(q^{\frac n2}a^{2m}\omega^{2\zeta m};q){\displaystyle\prod_{m=-\frac n2+1}^{\frac n2-1}}(q^{\frac n2+1}a^{2m}\omega^{2\zeta m};q)}.
\end{align}
We see that it is indeed identical to the twisted character of the universal $\mathcal{W}_{\infty}^{s,s}$ algebra \cite{Bonetti:2025kan} for general $c$. 
This analysis justifies our identification of the NIS action on the VOA as an outer-automorphism.

\section{Fusion rules and the charged index}
\label{sec:fusion}
In $\mathcal{N}=4$ $SU(N)$ SYM, we propose that the supersymmetric duality defect ($\mathcal{D}_d$) satisfies the fusion rules which are an extension of the ones in \cite{Kaidi:2021xfk,Choi:2021kmx,Kaidi:2022uux}\footnote{The second equation is slightly different from the one discussed in \cite{Kaidi:2021xfk,Choi:2021kmx,Kaidi:2022uux}. The difference is the part from the center of $SL(2,\mathbb{Z})$.}:
\begin{align}
\mathcal{D}_d \times \bar{\mathcal{D}}_d 
 & =   C_0, 
 \label{fusion2} \\
\mathcal{D}_d \times \mathcal{D}_d
 & =  (\mathcal{C} \tilde{\mathcal{R}}_0) \, C_0,
 \label{fusion1}
\end{align}
where 
\begin{align}
    C_0 = \frac{1}{|H^0(M_3, \mathbb{Z} )|}\sum_{\Sigma \in H_2(M_3, \mathbb{Z}_N)} L(\Sigma) \ , 
\end{align} 
is the condensation defect and $\bar{\mathcal{D}}_d$ is the orientation reversal of $\mathcal{D}_d$. In \eqref{fusion1}, $\mathcal{C}$ denotes the charge conjugation, which acts on local operators with the gauge charge as well as the background 2-form gauge field $B$ for the 1-form symmetry. The latter comes from the fact that $\sigma^2$ acts on the background 2-form gauge field as $B\rightarrow -B$ \cite{Kaidi:2021xfk, Choi:2021kmx}.  Also, $\tilde{\mathcal{R}}_0$ is the combined action of the center of $SU(4)_R$, whose action is given by $g_{r_0} = e^{-\frac{\pi i}{2}} {\bf 1} \in SU(4)_R$, and $(\mathcal{R}_\zeta)^2$ (giving the phase $\omega^{2 R_\zeta} = (-1)^{R_\zeta}$), which is necessary to preserve $\mathcal{N}=2$ supersymmetry parameterized by $\zeta$. 

Likewise, the triality defect satisfies the following fusion rules: 
\begin{align}
\bar{\mathcal{D}}_t \times \mathcal{D}_t
 &=     C_0, 
 \label{fusion3}\\
\mathcal{D}_t \times \mathcal{D}_t \times \mathcal{D}_t
 & =    (\mathcal{C} \tilde{\mathcal{R}}_0) \, \mathcal{Z}_{U(1)_N}  C_0,
 \label{fusion4}
\end{align}
The local part of the right-hand side of \eqref{fusion4} is the same as the duality defect case, since $(ST)^3= S^2$. The global part has the insertion of a $3d$ $U(1)$ Chern-Simons theory at level $N$ on the defect.

We test these fusion rules at the Schur index level. The insertion of the condensation defect $C_0$ does not change the index, since $H_2(S^3;\mathbb{Z})=\{ 0\} $. Thus, \eqref{fusion2} and \eqref{fusion3} say that the orientation reversal of the duality defect is actually its inverse at the index level. This is indeed the case since the orientation reversal acts on the ``fugacity" as $\omega \rightarrow \omega^{-1}$. 

The second fusion rule \eqref{fusion1} is shown by matching two independent computations. The left-hand side is obtained from the VOA twisted character, where the ``fugacity" is fixed as $\omega\rightarrow\omega^{2}=-1$, and the right-hand side is obtained from the $4d$ gauge theory computation from the outer-automorphism twisted or the charged index \cite{Zwiebel:2011wa,Mekareeya:2012tn}. Let us remark here that the left-hand side corresponds to the $U(1)_Y$ action with $\phi = \pi/2$. 

For the fusion of the triality defects \eqref{fusion4}, the check goes through as well. On the VOA side, we set the ``fugacity" as $\omega \rightarrow \omega^3 = -1$. On the gauge theory side, the computation is the same as the duality defect case since the $\mathcal{Z}_{U(1)_N}$ part does not act on the local operators. 
In the following, we focus on the fusion rule \eqref{fusion1} and demonstrate the detailed index computations on both sides.

\paragraph{Fusion from twisted character}
On the VOA side, the fusion is very simple to realize. First, compute the NIS-twisted character and then replace $\omega \rightarrow \omega^2$, implementing the action of $\mathcal{D}_d$ twice. Then, we set $\omega=e^{\frac{2\pi i}{4}}=i$ at the end for the proper action of $\mathcal{D}_d$, instead of ill-defined $U(1)_Y$ twisting. 

Let us explicitly show the calculation for the case of $N=2, 3, 4$. The spectra of $\mathfrak{psl}(2|2)$ primaries up to $h=4$ are given in Table \ref{tab:pslprimaries}. From this we can directly compute the twisted character. The expression for the $\chi_{\mathcal{V}(A_1)}$ was already given in \eqref{chiA1}. Similarly, for $N=3, 4$, we have
\begin{align}
\begin{split}
\chi_{\mathcal{V}(A_{2})}
&= 1 + \chi_{1}q + (\chi_{\frac32}-(\omega+\omega^{-1})\chi_{\frac12})q^{\frac{3}{2}} + (\chi_{2}+\chi_{1}+2-(\omega+\omega^{-1})\chi_{1}) q^2  \\
&~+ (\chi_{\frac52}+2\chi_{\frac32}+\chi_{\frac12}-(\omega+\omega^{-1})(\chi_{\frac32}+2\chi_{\frac12})) q^{\frac{5}{2}}  \\
&~+ (2\chi_{3}+(1-2\omega-2\omega^{-1})\chi_{2} +(5-2\omega-2\omega^{-1})\chi_{1}+3+\omega^{2}+\omega^{-2} ) q^3 
\label{chiA2}  \\
&~ + \Big( \chi_{\frac72}+(3-2\omega-2\omega^{-1})\chi_{\frac52}+(7+\omega^{2}-3\omega-3\omega^{-1}+\omega^{-2})\chi_{\frac32}  \\
& \qquad \quad +(3-5\omega-5\omega^{-1})\chi_{\frac12}\Big) q^{\frac{7}{2}} \\ 
&~ + \Big( 2\chi_{4}+(3-2\omega-2\omega^{-1})\chi_{3}  +(8-6\omega-6\omega^{-1})\chi_{2} 
   \\
& \qquad \quad +(10+2\omega^{2}-6\omega-6\omega^{-1}+2\omega^{-2})\chi_{1} +9+\omega^{2}-\omega-\omega^{-1}+\omega^{-2}\Big)q^4 \\
&~ + \ldots \ , 
\end{split} 
\end{align}
\begin{align}
\begin{split}
\chi_{\mathcal{V}(A_{3})}
&=1+ \chi_{1}q + (\chi_{\frac32}-(\omega+\omega^{-1})\chi_{\frac12})q^{\frac{3}{2}} + (2\chi_{2}+\chi_{1}+2-(\omega+\omega^{-1})\chi_{1}) q^2  \\
&~+ \Big(\chi_{\frac52}+2\chi_{\frac32}+2\chi_{\frac12}-(\omega+\omega^{-1})(2\chi_{\frac32}+2\chi_{\frac12})\Big) q^{\frac{5}{2}}  \\
&~+ \Big(3\chi_{3}+(3-2\omega-2\omega^{-1})\chi_{2} +(7-3\omega-3\omega^{-1})\chi_{1}+3+\omega^{2}+\omega^{-2} \Big) q^3 
 \\
&~ + \Big( 2\chi_{\frac72}+(4-4\omega-4\omega^{-1})\chi_{\frac52}+(9+\omega^{2}-6\omega-6\omega^{-1}+\omega^{-2})\chi_{\frac32}  \\
& \qquad \quad +(7+\omega^{2}-6\omega-6\omega^{-1}+\omega^{-2})\chi_{\frac12}\Big) q^{\frac{7}{2}} \label{chiA3} \\
&~+ \Big( 4\chi_{4}+(6-4\omega-4\omega^{-1})\chi_{3}+(18+\omega^{2}-8\omega-8\omega^{-1}+\omega^{-2})\chi_{2}  \\
&~\quad +(17+3\omega^{2}-11\omega-11\omega^{-1}+3\omega^{-2})\chi_{1} +12+\omega^{2}-4\omega-4\omega^{-1}+\omega^{-2}\Big)q^4 \\
&~+ \ldots \ . 
\end{split}  
\end{align}
Here we used shorthand notation $\chi_{j}$ for the spin-$j$ $\mathfrak{sl}(2)$ character with variable $a\omega^{\zeta}$. We simply plug in $\omega\rightarrow-1$ to obtain the `double' NIS-twisted character, which corresponds to inserting $\mathcal{D}_d\times \mathcal{D}_d$.

\paragraph{Gauge charge conjugation}
The right-hand side \eqref{fusion1} is computed from the $4d$ gauge theory. The charge conjugation for the gauge group acts on a field $\phi$ in the adjoint representation as $\phi \rightarrow - \phi^T$. Consider the superconformal index with the charge conjugation operator $\mathcal{C}$ inserted. This ``charged" or outer-automorphism twisted index should pick up the $\mathcal{C}$-invariant, gauge-invariant states. This results in effectively changing the $SU(N)$ gauge group to $G(N)$, where $G(N)=SO(N+1)$ for even $N$ and $Sp(N-1)$ for odd $N$, which are obtained by `folding' the Lie algebra of $SU(N)$ by the $\mathbb{Z}_2$ automorphism. There also exists a ``twisted-sector" which modifies the single-letter index part. This gives \cite{Zwiebel:2011wa, Mekareeya:2012tn}
\begin{align}
\mathcal{I}_{{\rm C}}(q, a)
&=  \int [dU_{G(N)}] \exp \left[ \sum_{m=1}^\infty \left( \frac{i(q^m,a^m)}{m} V_m + \frac{i(q^{2m},a^{2m})}{m} W_m \right)  \right],
\end{align}
where $[dU_{G(N)}]$ is the Haar measure of $G(N)$ group and
\begin{align}
V_m
&=     \sum_{i=1}^{[N/2]}(z_i^m+z_i^{-m})-(-1)^m \delta^+_N, ~~~ \delta^\pm_N = \frac{1}{2} (1\pm (-1)^N), \\
W_m
&=     \sum_{1\leq i<j\leq[N/2]} (z_i^m+z_i^{-m})(z_j^m+z_j^{-m}) + (-1)^m \delta^-_N V_m + [N/2].
\end{align}
The symbol $[x]$ denotes the greatest integer less than or equal to $x$. For $N=2,3$, $G(2)=SO(3)$ and $G(3)=Sp(2)$, the Haar measure $[dU_{G(N)}]$ is given by
\begin{align}
    [dU_{SO(3)}]&=-\frac{1}{2}\frac{dz}{2\pi iz}\left(z+\frac1z-2\right),\\
    [dU_{Sp(2)}]&=-\frac{1}{2}\frac{dz}{2\pi iz}\left(z^{2}+\frac{1}{z^{2}}-2\right).
\end{align}

On the right-hand side of \eqref{fusion1}, we have the additional $\tilde{\mathcal{R}}_0$ action. The actions of the central element $g_{r_0}=e^{-\frac{\pi i}{2}} {\bf 1} = (-1)^{r_0}$ on the elementary fields and the supercharges are given as
\begin{align}
F \rightarrow F, ~~~
\lambda_I \rightarrow i \lambda_I,~~~
(X,Y,Z) \rightarrow - (X,Y,Z), ~~~
\mathcal{Q}^I \rightarrow -i \mathcal{Q}^I. 
\end{align}
We have listed the charges of the Schur letters under $r_0$ in Table \ref{tab:chargesN=4}. The phase rotation of supercharges, $\mathcal{Q}^{1,2}$, can be canceled by $(\mathcal{R}_\zeta)^2$ action. Then the charged index ($\mathcal{C}\tilde{\mathcal{R}}_0$-twisted index) is 
\begin{align}  \label{chargedindex}
\mathcal{I}_{{\rm C}}(q,a;\zeta)
 =     {\rm Tr}_C (-1)^F q^{E - R} a^f \eta^{-r_0 + R_\zeta}|_{\eta=-1},
\end{align}
where the fugacity $\eta$ is introduced to keep track of the $\hat{\mathcal{R}}_0$ action which is set to $-1$ in the end. The single-letter Schur index twisted by $\tilde{R}_0$ is given by
\begin{align}
i_{\tilde{R}_0}(q, a; \eta; \zeta)
     = \frac{- 2q + q^{\frac{1}{2}}(a \eta^{-1+\zeta} + \frac{1}{a \eta^{1+\zeta}}) }{1-q}.
\end{align}
All in all, the Schur index of $SU(N)$ SYM representing the right hand side of \eqref{fusion1} is
\begin{align}
\mathcal{I}_{{\rm C},N} (q, a; \zeta)
 =     \int [dU_{G(N)}] \exp \left[ \sum_{m=1}^\infty \left( \frac{i_{\tilde{R}_0}(q^m,a^m,\eta^m)}{m} V_m + \frac{i_{\tilde{R}_0}(q^{2m},a^{2m},\eta^{2m})}{m} W_m \right)  \right] \Bigg|_{\eta=-1}.\label{chargedindex_int}
\end{align}

Now, let us compare the charge conjugation index with the VOA twisted character. The charged (outer-autmorphism twisted) indices for $SU(2)$, $SU(3)$ and $SU(4)$ SYM as $q$-series are computed from \eqref{chargedindex_int} to give
\begin{align}
\begin{split}
\mathcal{I}_{{\rm C},2}(q, a; \zeta)
&=   1 + \chi_1 q + 2 \chi_{\frac{1}{2}} q^{\frac32} + (\chi_2 + \chi_1 +1)q^2 + (2 \chi_{\frac32} + 2 \chi_{\frac{1}{2}}) q^{\frac52}  \\
&~+ (\chi_3+\chi_2+3\chi_1+1) q^3 + (2 \chi_{\frac52}+4\chi_{\frac32} +4\chi_{\frac{1}{2}}) q^{\frac72} \\
&~  + (\chi_4+\chi_3+4\chi_2+7\chi_1+3) q^4 + \ldots, 
\end{split} \\
\begin{split}
\mathcal{I}_{{\rm C},3}(q, a; \zeta)
&=   1 + \chi_1 q + (\chi_{\frac32}+2\chi_{\frac{1}{2}})q^{\frac32} + (\chi_2+3\chi_1+2) q^2 + (\chi_{\frac52}+4\chi_{\frac32}+5\chi_{\frac{1}{2}}) q^{\frac52}  \\
&~ + (2\chi_3+5\chi_2+9\chi_1+5) q^3 + (\chi_{\frac72}+7\chi_{\frac52}+15\chi_{\frac32}+13\chi_\frac{1}{2}) q^{\frac72} \\
&~ + (2\chi_4 + 7 \chi_3 + 20 \chi_2 + 26 \chi_1 + 13) q^4 + \ldots,
\end{split} \\
\begin{split}
\mathcal{I}_{{\rm C},4}(q, a; \zeta)
&=   1 + \chi_1 q + (\chi_{\frac32}+2\chi_{\frac{1}{2}})q^{\frac32} + (2\chi_2+3\chi_1+2) q^2 \\
&~ + (\chi_{\frac52}+6\chi_{\frac32}+6\chi_{\frac{1}{2}}) q^{\frac52} + (3\chi_3+7\chi_2+13\chi_1+7) q^3 \\
&~  + (2\chi_{\frac72}+12\chi_{\frac52}+23\chi_{\frac32}+21\chi_\frac{1}{2}) q^{\frac72} \\
&~ + (4\chi_4 + 14 \chi_3 + 36 \chi_2 + 45 \chi_1 + 22) q^4 + \ldots,
\end{split}
\end{align}
where we once again omitted the argument of the characters $\chi_j(a(-1)^\zeta)$. These expressions match with the twisted characters \eqref{chiA1} for $A_1$ and \eqref{chiA2}-\eqref{chiA3} for $A_2$ and $A_3$ for general $\zeta$, after taking $\omega\rightarrow-1$.

We performed the same check up to $N=8$, where the spectrum of $\mathfrak{psl}(2|2)$ primaries saturates that of $\mathcal{W}_{\infty}^{s,s}$ for $h\leq4$. We verified that the twisted character $\chi_{\mathcal{V}(A_{N-1})}(q,a;-1;\zeta)$ for the VOA and the charged index $\mathcal{I}_{{\rm C},N}(q,a;\zeta)$ for $SU(N)$ SYM agrees up to $q^{4}$ order, for all $N\leq8$. Therefore, the twisted character of the VOA \eqref{VOAtwcharacter} correctly captures the fusion rule \eqref{fusion1}, and the interpretation of the outer-automorphism-twisted character as the NIS-twisted index is consistent.

\section{Macdonald index} \label{sec:macdonald}
 In this section, we promote our previous discussion to the Macdonald index, which is a refined index that counts the same set of operators as the Schur index. The Macdonald index is defined as 
\begin{align}
    \mathcal{I}_{M}(q,a,T)={\rm{Tr}}_{M}(-1)^{F}q^{E-2R+r}a^{f}t^{R-r}={\rm{Tr}}_{M}(-1)^{F}q^{E-R}a^{f}T^{R-r},
\end{align} 
where $t=qT$ and the trace is taken over the 1/8-BPS states with $E - 2R -2 j_2 + r = 0$ and $j_1 - j_2 + r = 0$.\footnote{For the Schur index, the trace can be taken over the entire Hilbert space. However, this is not the case for the Macdonald index, and we should restrict to the slice with these conditions.} The Macdonald index is reduced to the Schur index when $t\rightarrow q$ (or $T\rightarrow 1$). The Macdonald index with the insertion of non-invertible symmetry defect $\mathcal{D}_{d,t}$ is given by
\begin{align}
    \mathcal{I}_{M,\mathcal{D}_{d,t}}(q,a,T;\omega;\zeta)={\rm Tr}_M (-1)^{F}q^{E-R}a^{f}T^{R-r}\omega^{s+R_{\zeta}}.
    \label{NISMacdonaldindex}
\end{align}

We propose that the defect Macdonald index is computed by the VOA twisted character refined with additional weight. This is possible due to the universal filtration that is present in any VOA, from which we can compute the `refined
 character' \cite{Song:2016yfd} that matches the Macdonald index. This relation has been further investigated in \cite{Fluder:2017oxm, Agarwal:2018zqi, Bonetti:2018fqz}, for example. 
Let us assign $\mathfrak{R}$-weight for each strong generator of $\mathcal{W}_{\infty}^{s,s}$, as table \ref{tab:Rweight}.
\begin{table}[h]
    \centering
    \begin{tabular}{|c||c|c|c|c|c|c|c|c|}
        \hline
        $\mathfrak{psl}(2|2)$ multiplet & \multicolumn{4}{|c|}{$\mathbb{J}$} & \multicolumn{4}{|c|}{$\mathbb{W}_{p>2}$} \\\hline
        Operators & $J$ & $G$ & $\tilde{G}$ & $T$ & $W_{p}$ & $G_{W_{p}}$ & $\tilde{G}_{W_{p}}$ & $T_{W_{p}}$ \\\hline
        $\mathfrak{R}$-weight & $1$ & $1$ & $1$ & $1$ & $\frac{p}{2}$ & $\frac{p}{2}$ & $\frac{p}{2}$ & $\frac{p}{2}$ \\\hline
    \end{tabular}
    \caption{$\mathfrak{R}$-weights of the strong generators of $\mathcal{W}_{\infty}^{s,s}$.}
    \label{tab:Rweight}
\end{table}
The $\mathfrak{R}$-filtration is an increasing sequence of subspaces of $\mathcal{W}_{\infty}^{s,s}$,
\begin{align}
    \mathfrak{F}_{\mathfrak{R}}=\text{span of states with $\mathfrak{R}$-weight $\mathfrak{R}-k$, $k\in\mathbb{Z}_{\geq0}$},~~~~ \mathfrak{F}_{\mathfrak{R}+1}\supseteq\mathfrak{F}_{\mathfrak{R}}.
\end{align}
It is claimed in \cite{Song:2016yfd, Agarwal:2018zqi, Bonetti:2025kan,Bonetti:2018fqz} that the $\mathfrak{R}$-filtration is identical with $R$-filtration from $4d$ $\mathcal{N}=4$ SYM for $\mathcal{W}_{\infty}^{s,s}$ and its truncation to $\mathcal{V}(A_{N-1})$. Therefore, the $\mathfrak{R}$-weight corresponds to the quantum number $R$ of $SU(2)_R$ in $4d$ superconformal theory. We propose that the defect Macdonald index \eqref{NISMacdonaldindex} is computed by the outer-automorphism twisted character of the VOA, refined by fugacity $T$:
\begin{align}
    \chi_{\mathcal{V}(A_{N-1})}(q,a,T;\omega;\zeta)={\rm Tr}(-1)^{F}q^{h}a^{j}T^{\mathfrak{R}-r}\omega^{-2r+\zeta j}.
\end{align}
Since the $\mathfrak{R}$-weight is identical for the states in the same $\mathfrak{psl}(2|2)$ multiplet, we can use the same formula for the contribution of the long and short $\mathfrak{psl}(2|2)$ multiplets \eqref{multiplets}-\eqref{multipletl0} with $\omega\rightarrow \omega T^{\frac12}$.

As a check for this proposal, we again consider the large-$N$ limit. From the supergravity multiplet structure listed in Table \ref{tab:kkschur}, the single-letter Macdonald index for the graviton multiplet is
\begin{align}
    i_{M}(q,a,T;\omega;\zeta)=\sum_{n=2}^{\infty}\frac{q^{\frac n2}T^{\frac n2}\chi_{\frac n2}-q^{\frac n2+\frac12}T^{\frac n2}(\omega T^{\frac12}+\omega^{-1}T^{-\frac12})\chi_{\frac n2-\frac12}+q^{\frac n2+1}T^{\frac n2}\chi_{\frac n2-1}}{1-q},
\end{align}
where we used shorthand notation $\chi_{j}=\chi_{j}(a\omega^{\zeta})$. Taking the Plethystic exponential, 
\begin{align}
    &\mathcal{I}_{M,\mathcal{D}_{d,t}}(q,a,T;\omega;\zeta)\nonumber\\
    &~~~~=\prod_{n=2}^{\infty}\frac{{\displaystyle\prod_{m=-\frac n2+\frac12}^{\frac n2-\frac12}(q^{\frac n2+\frac12}T^{\frac n2+\frac12}a^{2m}\omega^{1+2\zeta m};q)(q^{\frac n2+\frac12}T^{\frac n2-\frac12}a^{2m}\omega^{-1+2\zeta m};q)}}{{\displaystyle\prod_{m=-\frac n2}^{\frac n2}}(q^{\frac n2}T^{\frac n2}a^{2m}\omega^{2\zeta m};q){\displaystyle\prod_{m=-\frac n2+1}^{\frac n2-1}}(q^{\frac n2+1}T^{\frac n2}a^{2m}\omega^{2\zeta m};q)} \ , 
\end{align}
we get the full multi-particle index. 
Again, this is identical to the formula for the twisted character of $\mathcal{W}_{\infty}^{s,s}$, refined by $T$.

The defect Macdonald index computed as a twisted character is consistent with the fusion rule \eqref{fusion1} and \eqref{fusion4} for finite $N$. Let us demonstrate this explicitly for $N=2$ and $N=3$. The spectrum of $\mathfrak{psl}(2|2)$ primaries, dressed by the fugacity $T$ can be easily read off, which is presented in Table \ref{tab:Macdonaldpslprimaries} \cite{Bonetti:2025kan}. 
\begin{table}[t]
    {\small
    \begin{align}
    &\begin{tabular}{|c||c||c||c|c||c|c||c|c|}
    \hline
      $h$ & $1$ & $\frac32$ & \multicolumn{2}{|c||}{$2$} & \multicolumn{2}{|c||}{$\frac52$} & \multicolumn{2}{|c|}{$3$}\\\hline
      $j$ & $1$ & $\frac32$ & $2$ & $0$ & $\frac52$ & $\frac32$ & $3$ & $1$\\\hline\hline
      $N=2$ & $T$ & 0 & $T^{2}$ & 0 & 0 & 0 & $T^{3}$ & $T^{2}$ \\\hline
      $N=3$ & $~T^{2}~$ & $~~T^{\frac32}~~$ & $~T^{2}~$ & $~T^{2}~$ & $~~T^{\frac52}~~$ & $~~T^{\frac52}~~$ & $~~2T^{3}~~$ & $~~~T^{2}+T^{3}~~~$\\\hline
    \end{tabular}\nonumber\\
    &\begin{tabular}{|c||c|c|c|c||c|c|c|c|c|}
    \hline
      $h$ & \multicolumn{4}{|c||}{$\frac72$} & \multicolumn{5}{|c|}{$4$}\\\hline
      $j$ & \multicolumn{1}{|c|}{$\frac72$}& \multicolumn{1}{|c|}{$\frac52$} & $\frac32$ & $\frac12$ & $4$ & \multicolumn{1}{|c|}{$3$} & $2$ & \multicolumn{1}{|c|}{$1$} & $0$\\\hline\hline
      $N=2$ & 0 & 0 & 0 & 0 & $T^{4}$ & 0 & $T^{3}$ & 0 & $T^{2}$ \\\hline
      $N=3$ & $T^{\frac72}$ & $T^{\frac72}$ & $T^{\frac52}+T^{\frac72}$ & $T^{\frac72}$ & $2T^{4}$ & $T^{4}$ & $2T^{3}+T^{4}$ & $T^{3}$ & $T^{2}+T^{3}+T^{4}$ \\\hline
    \end{tabular}\nonumber
    \end{align}}
    \caption{Spectra of $\mathfrak{psl}(2|2)$ primaries of $\mathcal{V}(A_{N-1})$, when $N=2, 3$. Each entry is $\sum_{r}n_{h,j,r}(-1)^{F}T^{\mathfrak{R}-r}\omega^{-2r}$ with fixed $h$ and $j$.}
    \label{tab:Macdonaldpslprimaries}
\end{table}
For $SU(2)$ and $SU(3)$, the refined twisted characters are given as
\begin{align} \label{eq:twistedchMacsu2}
\begin{split}
    \chi_{\mathcal{V}(A_{1})}&= 1 + T\chi_1 q- (T^{\frac12}\omega^{-1}+T^{\frac32}\omega)\chi_{\frac12}q^{\frac32}+(T+T\chi_1+T^2\chi_2)q^2  \\
    &~ +\Big( (-T^{\frac12}\omega^{-1}-T^{\frac32}\omega)\chi_{\frac12}+(-T^{\frac32}\omega^{-1}-T^{\frac52}\omega )\chi_{\frac32}\Big)q^{\frac52}\\
    &~+\Big(T+(T+2T^2)\chi_1+T^2\chi_2+T^3\chi_3\Big)q^3  \\
    &~ +\Big((-T^{\frac12}\omega^{-1}-T^\frac32\omega-T^\frac32\omega^{-1}-T^\frac52\omega)\chi_{\frac12}+(-T^{\frac32}2\omega^{-1}-2T^\frac52\omega)\chi_{\frac32} \\
    &\qquad +(-T^{\frac52}\omega^{-1}-T^{\frac72}\omega)\chi_{\frac52}\Big)q^{\frac72} \\
    &~ +\Big(T+2T^2+(T+T\omega^{-2}+4T^2+T^3\omega^2)\chi_1 \\
    &\qquad + (2T^2+2T^3)\chi_2+T^3\chi_3+T^4\chi_4\Big)q^4+\ldots,
\end{split}
\end{align}
\begin{align} \label{eq:twistedchMacsu3}
\begin{split}
    \chi_{\mathcal{V}(A_{2})}&= 1 + T\chi_1 q+ \Big((-T^{\frac12}\omega^{-1}-T^{\frac32}\omega)\chi_{\frac12}+T^\frac32\chi_\frac32\Big)q^{\frac32} \\
    &~+\Big(T+T^{2}+(T-T\omega^{-1}-T^{2}\omega)\chi_1+T^2\chi_2\Big)q^2  \\
    &~ +\Big( (-T^{\frac12}\omega^{-1}+T^\frac32-T^{\frac32}\omega^{-1}-T^{\frac32}\omega-T^\frac52\omega)\chi_{\frac12} \\
    &~~~~+(T^\frac32-T^\frac32\omega^{-1}+T^\frac52-T^\frac52\omega)\chi_{\frac32}+T^\frac52\chi_\frac52\Big)q^{\frac52}  \\
    &~ +\Big(T+T\omega^{-2}+2T^2+T^3\omega^{2}+(T-T\omega^{-1}+3T^2-T^{2}\omega^{-1}-T^{2}\omega+T^3-T^3\omega)\chi_1 \\
    &~~~~+(T^2-2T^2\omega^{-1}-2T^3\omega)\chi_2+2T^3\chi_3\Big)q^3\\
    &~+\Big((-T^{\frac12}\omega^{-1}+T^{\frac32}1-3T^\frac32\omega^{-1}-T^\frac32\omega+2T^\frac52-T^\frac52\omega^{-1}-3T^\frac52\omega-T^\frac72\omega)\chi_{\frac12}\\
    &~~~~ +(T^{\frac32}+T^{\frac32}\omega^{-2}-2T^{\frac32}\omega^{-1}+5T^\frac52-T^{\frac52}\omega^{-1}-2T^{\frac52}\omega+T^{\frac72}-T^{\frac72}\omega+T^{\frac72}\omega^2)\chi_{\frac32}  \\
    &~~~~ +(2T^{\frac52}-2T^{\frac52}\omega^{-1}+T^{\frac72}-2T^{\frac72}\omega)\chi_{\frac52}+T^{\frac72}\chi_\frac72\Big)q^{\frac72}  \\
    &~ +\Big(T+T\omega^{-2}+5T^2-T^2\omega^{-1}+2T^3-T^3\omega+T^3\omega^2+T^4+(T+T\omega^{-2}-T\omega^{-1}\\
    &~~~~ +5T^2+T^2\omega^{-2} -4T^2\omega^{-1}-T^2\omega +4T^3-T^3\omega^{-1}-4T^3\omega+T^3\omega^2-T^4\omega+T^4\omega^2)\chi_1 \\
    &~~~~+(2T^2-4T^2\omega^{-1} +5T^{3}-2T^3\omega^{-1} -4T^3\omega+T^{4}-2T^4\omega)\chi_2\\
    &~~~~+(2T^3-2T^3\omega^{-1}+T^{4}-2T^4\omega)\chi_3+2T^4\chi_4\Big)q^4+\ldots.
\end{split}
\end{align}

Now, the Macdonald index for the right-hand side of \eqref{fusion1} corresponds to the charge conjugation or outer-automorphsim twisted index. It can be computed using  \eqref{chargedindex_int}, with the Macdonald single-letter index,
\begin{align}
    i_{M,\tilde{R}_{0}}(q,a,T;\eta;\zeta)=\frac{-q(T+1)+q^{\frac12}T^{\frac12}(a \eta^{-1+\zeta} + \frac{1}{a \eta^{1+\zeta}})}{1-q}.
\end{align}
This gives for $SU(2)$ and $SU(3)$ SYM
\begin{align}
\begin{split}
    \mathcal{I}_{{\rm C},2}&=1+T\chi_1 q + (T^{\frac12}+T^{\frac32})\chi_{\frac12}q^{\frac32} +(T+T\chi_1+T^2\chi_2)q^{2} \\
    &~ +\Big((T^{\frac12}+T^{\frac32})\chi_{\frac12}+(T^{\frac32}+T^{\frac52})\chi_{\frac32}\Big)q^{\frac52}\\
    &~+\Big(T+(T+2T^2)\chi_1+T^2\chi_2+T^3\chi_3\Big)q^3  \\
    &~ +\Big((T^{\frac12}+2T^{\frac32}+T^{\frac52})\chi_{\frac12}+(2T^{\frac32}+2T^{\frac52})\chi_{\frac32}+(T^{\frac52}+T^{\frac72})\chi_{\frac52}\Big)q^\frac72\\
    &~ +\Big(T+2T^2+(2T+4T^2+T^3)\chi_1+(2T^2+2T^3)\chi_2+T^3\chi_3+T^4\chi_4\Big)q^4 \\
    &~+\ldots,
\end{split}
\end{align}
\begin{align}
\begin{split}
    \mathcal{I}_{{\rm C},3}&=1+T\chi_1 q + \Big((T^{\frac12}+T^{\frac32})\chi_{\frac12}+T^\frac32\chi_\frac32\Big)q^{\frac32} \\
    &~+\Big(T+T^2+(2T+T^2)\chi_1+T^2\chi_2\Big)q^{2} \\
    &~ +\Big((T^{\frac12}+3T^{\frac32}+T^\frac52)\chi_{\frac12}+(2T^{\frac32}+2T^{\frac52})\chi_{\frac32}+T^\frac52\chi_\frac52\Big)q^{\frac52}  \\
    &~+\Big(2T+2T^2+T^3+(2T+5T^2+2T^3)\chi_1+(3T^2+2T^3)\chi_2+2T^3\chi_3\Big)q^3  \\
    &~ +\Big((T^{\frac12}+5T^{\frac32}+6T^{\frac52}+T^\frac72)\chi_{\frac12}+(4T^{\frac32}+8T^{\frac52}+3T^\frac72)\chi_{\frac32}+ \\
    & \qquad \quad  +  (4T^{\frac52}+3T^{\frac72})\chi_{\frac52}+T^\frac72\chi_\frac72\Big)q^\frac72\\
    &~ +\Big(2T+6T^2+4T^3+T^4+(3T+11T^2+10T^3+2T^4)\chi_1 \\
    & \qquad \quad +(6T^2+11T^3+3T^4)\chi_2+(4T^3+3T^4)\chi_3+2T^4\chi_4\Big)q^4+\ldots \ .
\end{split}
\end{align}
They agree with the twisted characters \eqref{eq:twistedchMacsu2} and \eqref{eq:twistedchMacsu3} when we fix $\omega\rightarrow-1$, as expected. 

As we have done for the case of the Schur index, we have checked that the $\chi_{\mathcal{V}(A_{N-1})}$ at $\omega\rightarrow-1$ and $\mathcal{I}_{{\rm C},N}$ agree as a $q$-series up to $q^4$, for $N\leq8$.

\section{Conclusions and discussions}
\label{sec:conclusion}
In this paper, we propose that the duality non-invertible symmetry acts as an outer automorphism on the Schur sector described by a vertex operator algebra. We identified the NIS-twisted Schur (and Macdonald) index of $\mathcal{N}=4$ $SU(N)$ SYM theory as the outer-automorphism twisted character of the associated VOA $\mathcal{V}(A_{N-1})$ (with $\mathfrak{R}$-filtration). In the SCFT/VOA correspondence, it was already suggested that line and surface defects (which are codimension-two or three, and are not topological) are associated with the non-trivial modules of VOA or their combinations \cite{Cordova:2016uwk, Cordova:2017ohl}. 
What we find here extends the SCFT/VOA dictionary by incorporating the topological defect. We confirmed that the proposed index satisfies the defect fusion rules. Furthermore, we checked that the large $N$ limit of the index, obtained from the twisted character of $\mathcal{W}_\infty^{s,s}$ (without any truncation), matches the twisted index computed from supergravity. 

The superconformal index, which is the partition function on $S^1 \times S^3$, does not capture the global structure of the gauge theory. This washes out the most salient feature of the non-invertible symmetry defect, which acts on the higher-dimensional operators.  The fusion rule simply reduces to the invertible one on $S^1\times S^3$ spacetime. To see the effect of non-invertibility, we need to consider topologically non-trivial spacetime. The lens space index, which is a partition function on $S^1 \times L(p, 1)$ \cite{Benini:2011nc}, will be the most useful and tractable arena to study. It was found in \cite{Fluder:2017oxm} that the lens index reproduces the character of the twisted module of the associated VOA. It would be interesting to consider the outer-automorphism twist in this set-up, and see whether this is consistent with the properties of the NIS defect. The twisted VOA, along with the NIS, forms a more elaborate algebraic structure, known as the chiral tube algebra \cite{Benjamin:2026oqw}. It would be interesting to further clarify its role in the SCFT/VOA correspondence.  

It is remarkable that the automorphism of the VOA is associated with the non-trivial defect in the $4d$ gauge theory. It is thus interesting to see whether this correspondence is generalized to other theories, especially $\mathcal{N}=2$ SCFTs. The non-invertible defects of $\mathcal{N}=2$ class $\mathcal{S}$ theories were classified in \cite{Bashmakov:2022uek,Antinucci:2022cdi} for the case without punctures. The associated VOAs to the class $\mathcal{S}$ theories were given in \cite{Beem:2014rza}. It is natural to conjecture that automorphisms of a VOA give (possibly a subset of) non-invertible defects in the class $\mathcal{S}$ theories. It is known that the $A_1$ theory for a genus 2 surface does have a non-trivial automorphism \cite{Kiyoshige:2020uqz, Beem:2021jnm}.

\begin{acknowledgments}
We thank Saghar S.~Hosseini, Emily Nardoni and Yuji Tachikawa for useful discussions and for collaborating at an early stage of this work. We especially thank Yuji Tachikawa for suggesting the project and helpful correspondences. 
We also thank Takahiro Nishinaka and Kantaro Ohmori for useful discussions and comments.
The work of H.M. and J.S.~is supported by the National Research Foundation of Korea (NRF) grants RS-2024-00405629 and RS-2026-25482546, and the KAIST-KIAS collaboration program. 
The work of J.S.~is also supported in part by the Walter Burke Institute for Theoretical Physics and the Leinweber Forum for Theoretical Physics at Caltech and by the U.S.~Department of Energy, Office of Science, Office of High Energy Physics, under Award Number DE-SC0011632.
\end{acknowledgments}

\appendix
\section{Conventions for Supersymmetry}
\label{sec:convention}
We denote the supercharges of $\mathcal{N}=4$ superconformal symmetry by $\mathcal{Q}_\alpha^I$, $\tilde{\mathcal{Q}}_{I \dot{\alpha}}$, $\mathcal{S}^\alpha_I$, $\tilde{\mathcal{S}}^{I \dot{\alpha}}$ ($I=1,2,3,4$). These form $\mathfrak{su}(2,2|4)$ superconformal algebra together with $\mathfrak{su}(4)_R$ generator $\mathcal{R}^I_{~J}$. The convention of the tilded spinor is
  \begin{equation}
  (\mathcal{Q}_\alpha^I)^* = \tilde{\mathcal{Q}}_{I \dot{\alpha}}.
  \end{equation}
The Hermitian conjugate is given by 
  \begin{equation}
  (\mathcal{Q}^I_\alpha)^\dagger
   =     \mathcal{S}_I^\alpha, ~~~
(\tilde{\mathcal{Q}}_{I\dot{\alpha}})^\dagger
   =     \tilde{\mathcal{S}}^{I \dot{\alpha}}.
  \end{equation}
The commuting relations of the supercharges $\mathcal{Q}_\alpha^I$ are given by
  \begin{equation}
  \{ \mathcal{Q}_\alpha^I, \tilde{\mathcal{Q}}_{J \dot{\beta}} \}
   =     2 \delta^I_J \mathcal{P}_{\alpha \dot{\alpha}}, ~~~
  \{ \mathcal{Q}_\alpha^I, \mathcal{S}_J^\beta \}
   =     2 \delta^I_J \delta^\beta_\alpha H + 4 \delta^I_J \mathcal{M}_\alpha^{~\beta} - 4 \delta^\beta_\alpha \mathcal{R}^I_{~J}
  \end{equation}
where $H$ is the Hamiltonian whose eigenvalue gives the scaling dimensions, and  $\mathcal{R}^I_{~J}$ is the generator of the $SU(4)_R$ symmetry. 

Now we focus on the subalgebra $\mathfrak{su}(2,2|2) \times \mathfrak{su}(2)_f \subset \mathfrak{su}(2,2|4)$. We denote the Cartan generators of the $\mathfrak{su}(2)_R \times \mathfrak{u}(1)_r $ of $\mathfrak{su}(2,2|2)$ and the $\mathfrak{su}(2)_f$ by
  \begin{equation}
  R
   =     \frac{1}{2} (\mathcal{R}^1_1 - \mathcal{R}^2_2), ~~~
  r 
   =     \mathcal{R}^1_1 + \mathcal{R}^2_2, ~~~
  F
   =     \mathcal{R}^1_1 + \mathcal{R}^2_2 + 2 \mathcal{R}^3_3.
  \end{equation}
The charges of the supercharges $\mathcal{Q}_\alpha^I$ under these are given in Table \ref{tab:supercharges}. Here we also list the charges under the following combinations which we will use for define the superconformal index and its Schur limit:
  \begin{align}
  \delta_1
   &=     \frac{1}{2} \{ \mathcal{Q}^1_-, \mathcal{S}_1^- \}
   =     E - 2j_1 -2 R - r, \\
   \delta_2
   &=     \frac{1}{2} \{ \tilde{\mathcal{Q}}_{2 \dot{-}}, \tilde{\mathcal{S}}^{2 \dot{-}} \}
   =     E - 2j_2 -2 R + r.
  \end{align}

\begin{table}
    \centering
    \begin{tabular}{|c||c|c|c|c|c||c|c|}
         \hline
         & $j_1$ & $j_2$ & $R$  & $r$  & $f$ & $\delta_1$ & $\delta_2$ \\
         \hline 
         $\mathcal{Q}^1_+$ & $\frac{1}{2}$ & $0$ & $\frac{1}{2}$ & $\frac{1}{2}$ & $0$ & & $0$ \\
         $\mathcal{Q}^1_-$ & $-\frac{1}{2}$ & $0$ & $\frac{1}{2}$ & $\frac{1}{2}$ & $0$ & $0$ & $0$ \\
         $\tilde{\mathcal{Q}}_{1 \dot{+}}$ & $0$ & $\frac{1}{2}$ & $-\frac{1}{2}$ & $-\frac{1}{2}$ & $0$ & & $0$ \\
         $\tilde{\mathcal{Q}}_{1 \dot{-}}$ & $0$ & $-\frac{1}{2}$ & $-\frac{1}{2}$ & $-\frac{1}{2}$ & $0$ & & \\
         \hline
         $\mathcal{Q}^2_+$ & $\frac{1}{2}$ & $0$ & $-\frac{1}{2}$ & $\frac{1}{2}$ & $0$ & $0$ & \\
         $\mathcal{Q}^2_-$ & $-\frac{1}{2}$ & $0$ & $-\frac{1}{2}$ & $\frac{1}{2}$ & $0$ &  & \\
         $\tilde{\mathcal{Q}}_{2 \dot{+}}$ & $0$ & $\frac{1}{2}$ & $\frac{1}{2}$ & $-\frac{1}{2}$ & $0$ & $0$ & \\
         $\tilde{\mathcal{Q}}_{2 \dot{-}}$ & $0$ & $-\frac{1}{2}$ & $\frac{1}{2}$ & $-\frac{1}{2}$ & $0$ & $0$ & $0$ \\
         \hline
         $\mathcal{Q}^3_+$ & $\frac{1}{2}$ & $0$ & $0$ & $-\frac{1}{2}$ & $1$ & $0$ & $0$ \\
         $\mathcal{Q}^3_-$ & $-\frac{1}{2}$ & $0$ & $0$ & $-\frac{1}{2}$ & $1$ &  & $0$ \\
         $\tilde{\mathcal{Q}}_{3 \dot{+}}$ & $0$ & $\frac{1}{2}$ & $0$ & $\frac{1}{2}$ & $-1$ & $0$ & $0$ \\
         $\tilde{\mathcal{Q}}_{3 \dot{-}}$ & $0$ & $-\frac{1}{2}$ & $0$ & $\frac{1}{2}$ & $-1$ & $0$ & \\
         \hline
         $\mathcal{Q}^4_+$ & $\frac{1}{2}$ & $0$ & $0$ & $-\frac{1}{2}$ & $-1$ & $0$ & $0$ \\
         $\mathcal{Q}^4_-$ & $-\frac{1}{2}$ & $0$ & $0$ & $-\frac{1}{2}$ & $-1$ &  & $0$ \\
         $\tilde{\mathcal{Q}}_{4 \dot{+}}$ & $0$ & $\frac{1}{2}$ & $0$ & $\frac{1}{2}$ & $1$ & $0$ & $0$ \\
         $\tilde{\mathcal{Q}}_{4 \dot{-}}$ & $0$ & $-\frac{1}{2}$ & $0$ & $\frac{1}{2}$ & $1$ & $0$ & \\
         \hline
    \end{tabular}
    \caption{Charges of $\mathcal{N}=4$ supercharges under the R-symmetry in $\mathcal{N}=2$ notations as well as $\delta_{1,2}$ \eqref{delta12}.}
    \label{tab:supercharges}
\end{table}

We also use the R-charges $(R_1, R_2, R_3)$ which are given in terms of the generators $\mathcal{R}^I_{J}$ by 
\begin{align}
 R_1 
  = \mathcal{R}^1_1 - \mathcal{R}^2_2, ~~~
 R_2 
 = \mathcal{R}^2_2 - \mathcal{R}^3_3, ~~~
 R_3 
 = \mathcal{R}^3_3 - \mathcal{R}^4_4,
\end{align}
with $\mathcal{R}^4_4 = - (\mathcal{R}^1_1+\mathcal{R}^2_2+\mathcal{R}^3_3)$. In terms of these charges, the $\mathcal{N}=2$ R-symmetry charges are written as
\begin{align}
R 
 =     \frac{R_1}{2}, ~~~
r
 =     \frac{R_1}{2} + R_2 + \frac{R_3}{2}, ~~~
F 
 =     R_3.
\end{align}
Under these $(R_1, R_2, R_3)$ that the supercharges have the following charges:
\begin{align}
& \mathcal{Q}^1: (1, 0, 0), \\
&\mathcal{Q}^2: (-1, 1, 0), \\
&\mathcal{Q}^3: (0, -1, 1), \\
&\mathcal{Q}^4: (0, 0, -1).
\end{align}

The on-shell supersymmetry transformations of the local fields are given by
  \begin{align}
  \delta \phi_{[IJ]}
  &=     \eta_{[ I}^\alpha \lambda_{J ] \alpha} + \epsilon_{IJKL} \bar{\eta}^{K}_{\dot{\alpha}} \bar{\lambda}^{L \dot{\alpha}}, \\
  \delta \lambda_{I \alpha}
  &=    \eta^{\beta}_I F_{\alpha \beta} + D_{\alpha \dot{\alpha}} \phi_{IJ} \bar{\eta}^{I \dot{\alpha}} + [ \phi_{IJ}, \phi^{JK} ] \eta_{K \alpha}, \\
  \delta A_{\alpha \dot{\alpha}}
  &=    i \eta_{I\alpha} \bar{\lambda}^I_{\dot{\alpha}} + i \bar{\eta}^I_{\dot{\alpha}} \lambda_{I \alpha}.
  \label{susytransf}
  \end{align}
where $\phi^{IJ} = \frac{1}{2} \epsilon^{IJKL} \phi_{KL}$.
We denote the scalar fields $\phi_{IJ}$ as three complex scalars:
  \begin{equation}
  \phi_{12} 
   =     \bar{X}, ~~~
  \phi_{24}
   =     Y, ~~~
  \phi_{23}
   =     Z.
  \end{equation}

\section{``$U(1)_Y$ twisted" index}
\label{sec:U1Y}
The ill-defined object which is closely related to the NIS twisted index is $\mathcal{Y}= \omega^{\frac{Y}{2}+R_\zeta}$ twisted index \eqref{YSchurindex}. As noted in Section \ref{sec:index}, this is not a symmetry, but we will formally compute it. 

It is clear from the Table \ref{tab:chargesN=4} that the single-letter index is 
\begin{align}
    i_{Y}(q, a; \omega; \zeta)
     = \frac{- q(\omega + \omega^{-1}) + q^{\frac{1}{2}}(a \omega^\zeta + \frac{1}{a \omega^\zeta}) }{1-q}.
    \label{YsingleSchur}
\end{align}
The index of $SU(2)$ SYM is given 
The twisted index \eqref{SchurNIS} for $G=SU(2)$ is given by
\begin{align}
    \mathcal{I}_{Y, SU(2)}(q, a; \omega; \zeta)
    &= \int [dU_{SU(2)}]{\rm PE}[i_{Y}(q, a, \omega,\zeta)\chi_{\rm adj}(z)] \nonumber \\
    &=\frac{1}{2} \oint \frac{dz}{2 \pi i z} (1-z^2)(1-z^{-2}) \frac{(z^{\pm2}\omega^{\pm}q;q)(\omega^{\pm}q;q)}{(\sqrt{q}z^{\pm 2}(a\omega^\zeta)^{\pm}; q)(\sqrt{q}(a\omega^\zeta)^{\pm}; q)},
    \label{exact}
\end{align}
where $(x;q)=\prod_{k=0}^\infty (1-xq^k)$ is the $q$-Pochhammer symbol and $(x^{\pm};q) = (x;q)(x^{-1};q)$. We compute it as the series expansion in $q$
\begin{align}
\begin{split}
\mathcal{I}_{Y, SU(2)} 
&= 1+\chi_1 q-(\omega+\omega^{-1})\chi_{\frac{1}{2}} q^{\frac{3}{2}}
  + (\chi_2+\chi_1 + 1 +(1-\omega)(1-\omega^{-1}))q^2  \\
&~~ - \left( (\omega+\omega^{-1})(\chi_{\frac{3}{2}}+\chi_{\frac{1}{2}})+(\omega+\omega^{-1})(1-\omega)(1-\omega^{-1})\chi_{\frac{1}{2}}\right)q^{\frac{5}{2}}  \\
&~~ + (\chi_3 + \chi_2 +3\chi_1 + 1 + (1-\omega)(1-\omega^{-1})(\chi_3 +(\omega+\omega^{-1})^2)) q^3 +\ldots.
\end{split}
\end{align}

The difference between the NIS twisted index \eqref{chiA1} and $\mathcal{Y}$-twisted index is always proportional to the factor $(1-\omega)(1-\omega^{-1})$, which does not vanish when $\omega=e^{\pi i /2}$ and $e^{\pi i/3}$. Let us track which $4d$ operators are responsible for this factor. For example, at the $q^2$ order, there are two operators $\mathcal{O}_1 ={\rm Tr}(\lambda_2 \bar{\lambda}^1)$ and $\mathcal{O}_2 ={\rm Tr}(Y Z \bar{\lambda}^1)$ which cancel in the untwisted index. The on-shell supersymmetry transformation \eqref{susytransf} is roughly $\mathcal{Q}^1 \lambda_2 = F + [Y, Z]$ for the supercharge $\mathcal{Q}^1$ defining the Schur index. Thus $\mathcal{Q}^1 \mathcal{O}_1 \sim \mathcal{O}_2$. The boson-fermion pair cancellation means these are not in the $\mathcal{Q}_1$-cohomology. With the $U(1)_Y$ twist, $\mathcal{Y}$ acts on $\lambda_2$ and $X$, $Z$ with different phases, because $U(1)_Y$ is not a symmetry of the equations of motion for a non-Abelian gauge theory. They give contributions $1$ and $\omega$ respectively at the $q^2$ order. 

Let $g_\mathcal{Y}$ be an action of $\mathcal{Y}$ on the Hilbert space, then the non-cancellation means $[g_\mathcal{Y} , \mathcal{Q}_1] \neq 0$, even though $\mathcal{Q}_1$ is invariant under $\mathcal{Y}$. For the NIS-twisted index, for which the action on the Hilbert space is $g_{\mathcal{D}}$, $[g_\mathcal{D}, \mathcal{Q}_1]=0$. This is needed for the pairs not in $\mathcal{Q}_1$-cohomology to cancel. The VOA twisted character automatically does this, or it does so after taking $\mathcal{Q}_1$-cohomology. Note that for the $U(1)$ gauge theory $\mathcal{Y}$ is a symmetry of the equations of motion; thus, there is no such non-cancellation.

\section{Defect index of $\mathcal{N}=4$ $U(1)$ gauge theory}
\label{sec:U1}
The non-invertible defect of the Maxwell theory at the special values of the coupling constant $\tau_{U(1)}$ is considered in \cite{Choi:2021kmx} with a slight distinction with the $SU(N)$ case. This is generalized to $\mathcal{N}=4$ $U(1)$ gauge theory by adding free scalars and fermions. (See \cite{Sela:2024okz} for the related discussion.)

The $U(1)$ gauge theory has the $1$-form symmetry $U(1)_e^{(1)}\times U(1)_m^{(1)}$. Gauging the $\mathbb{Z}_p$ subgroup of the $U(1)_e^{(1)}$ electric $1$-form symmetry rescales the gauge field as $A \rightarrow A/p$. This means the coupling is now $\tau_{U(1)}'=\tau_{U(1)}/p^2$. 

The $U(1)$ gauge theory also has electric-magnetic duality. This acts on the coupling constant as $\tau_{U(1)} \rightarrow - 1/\tau_{U(1)}$, and on the field strength as $(F, F_D)^T \rightarrow (-F_D, F)^T$, where $F_D=\frac{4 \pi}{e^2} \ast F + \frac{\theta}{2 \pi}F$. In terms of the (anti-)self-dual field strength $F^{\pm} = \frac{1}{2}(F \mp i \ast F)$, the dual ones are given by $F_D^{+} = \tau F^+$ and $F_D^{-} = \bar{\tau} F^-$. Thus, by the duality action, $F^+ \rightarrow - \tau F^+$ and $F^- \rightarrow - \bar{\tau} F^-$.

The theory with $\tau_{U(1)}=i$ has the duality defect which is invertible. On the other hand, the theory with $\tau_{U(1)} = i p$ with an integer $p>1$ has a non-invertible duality defect. The gauging of the $\mathbb{Z}_p$ $1$-form symmetry sends the coupling as $\tau = i p \rightarrow i/p$, then the electric-magnetic duality sends the coupling to the original value. This acts on the self-dual field strength as $F^+ \rightarrow F^+/p \rightarrow - i F^+$ and on the anti-self-dual oppositely. This is seen as a part of $U(1)_Y$ symmetry \eqref{U1Y} with phase $\phi=\pi/4$.

To make the defect supersymmetric, we combine this $U(1)_Y$ action with that of $\mathcal{R}_\zeta$ as in the $SU(N)$ SYM case. The charges of the local fields under $R_\zeta$ and $U(1)_Y$ are summarized in Table \ref{tab:chargesN=4}. 
 
The Schur index of this free theory is simply the Plethystic exponential of the letter index. Without any defect, it is given by
\begin{align}
\mathcal{I}_{U(1)}(q,a)
 =     {\rm PE}[i(q, a)]  
 =     \frac{(q;q)^2}{(\sqrt{q}a^{\pm};q)}.
\label{U(1)Schur}
\end{align}
The defect index of the theory with $\tau_{U(1)} = i p$ is obtained by taking the charges into account in the letter index
\begin{align}
\mathcal{I}_{U(1),p}(q, a; \omega; \zeta)
& =     \frac{(\omega^{\pm} q;q)}{(\sqrt{q}(a \omega^{\zeta})^{\pm};q)},
\end{align}
where $\omega=e^{\pi i/2}$. This does not depend on $p$. The actions of the defects on the Schur sector (even more the sector contributing to the full superconformal index) are exactly the same. Thus the Schur index cannot distinguish the non-invertibility of the defect.

Let us check that this index is compatible with the fusion rules. It was discussed in \cite{Choi:2021kmx,Choi:2022zal} that the fusion of this type of duality defects gives the similar defect as the non-Abelian case. We argue that this is exactly the same as the fusion rule for $SU(N)$ case \eqref{fusion2} and \eqref{fusion1}. This is checked at the index level: the gauge charge conjugation $\mathcal{C}$ acts on all the single letters as the sign flip in the $U(1)$ case, while the supercharges are invariant, and the charges under $r_0$ and $R_\zeta$ are listed in Table \ref{tab:chargesN=4}. Therefore, the $\mathcal{C}\tilde{\mathcal{R}}_0$ - twisted index that corresponds to the right-hand side of the fusion rule \eqref{fusion1} is given by
\begin{align}
\mathcal{I}_{U(1), C}
 =    \frac{(\eta^{\pm} q;q)}{(\sqrt{q}(a \eta^{\zeta})^{\pm};q)},
\end{align}
where $\eta=-1$. This coincides with $\mathcal{I}_{U(1),p}(q, a; \omega^2=-1; \zeta)$ that corresponds to the left-hand side of \eqref{fusion1}.

\bibliographystyle{JHEP}
\bibliography{ref}

\end{document}